\documentclass[fleqn,usenatbib]{mnras}

\usepackage{newtxtext,newtxmath}

\usepackage[T1]{fontenc}
\usepackage{ae,aecompl}

\usepackage{graphicx}	
\usepackage{amsmath}	
\usepackage{amssymb}	
\usepackage[english]{babel}    
\usepackage{subfig}
\usepackage{array, booktabs, makecell}
\usepackage{siunitx}
\usepackage[version=4]{mhchem}

\title[Environments of Radio-WISE Selected Galaxies]{The Environments of Luminous Radio - WISE Selected Infrared Galaxies}

\author[J.I. Penney et al.]{
J. I. Penney,$^{1}$\thanks{E-mail: jip3@le.ac.uk}
A. W. Blain,$^{1}$
D. Wylezalek,$^{2}$
N. A. Hatch,$^{3}$
C. Lonsdale,$^{4}$
\and A. Kimball,$^{5}$
R. J. Assef,$^{6}$
J.J. Condon,$^{4}$
P. R. M. Eisenhardt,$^{7}$
S. F. Jones,$^{8}$
\and M. Kim,$^{9,10}$
M. Lacy,$^{4}$
S. I. Muldrew,$^{1}$
S. Petty,$^{11}$
A. Sajina,$^{12}$
A. Silva,$^{12,13}$
\and D. Stern,$^{7}$
T. Diaz-Santos,$^{6}$
C-W. Tsai$^{14},$
J. Wu$^{14,15}$
\\
$^{1}$University of Leicester, Department of Physics and Astronomy, University Road, LE1 7RH, UK \\
$^{2}$Department of Physics and Astronomy, Johns Hopkins University, Baltimore, MD 21218, USA\\
$^{3}$School of Physics and Astronomy, University of Nottingham, University Park, NG7 2RD, UK\\
$^{4}$National Radio Astronomy Observatory, 520 Edgemont Road, Charlottesville, VA 22903, USA\\
$^{5}$National Radio Astronomy Observatory, 1003 Lopezville Road, Socorro, NM 87801, USA\\
$^{6}$N{\'u}cleo de Astronom{\'i}a de la Facultad de Ingenier{\'i}a y Ciencias, Universidad Diego Portales, Av. Ej{\'e}rcito Libertador 441, Santiago, Chile\\
$^{7}$Jet Propulsion Laboratory, California Institute of Technology, 4800 Oak Grove Drive, CA 91109, USA\\
$^{8}$Department of Space, Earth, and Environment, Chalmers University of Technology, Onsala Space Observatory, SE-43992, Onsala, Sweden\\
$^{9}$Korea Astronomy and Space Science Institute, Daejeon 305-348, Korea\\
$^{10}$Department of Astronomy and Atmospheric Sciences, Kyungpook National University, Daegu 702-701, Korea \\
$^{11}$NorthWest Research Associates, 4118 148th Ave NE, Redmond, WA 98052, USA\\
$^{12}$Department of Physics and Astronomy, Tufts University, 574 Boston Ave, MA 02155, USA\\
$^{13}$Observatory of Japan, National Institutes of Natural Sciences, 2-21-1 Osawa, Mitaka, Tokyo 181-8588, Japan\\
$^{14}$Department of Physics and Astronomy, University of California, Los Angeles, CA 90095, USA\\
$^{15}$National Astronomical Observatories, Chinese Academy of Sciences, 20A Datun Road, Chaoyang District, Beijing, 100012, China
}

\author[J.I. Penney et al.]{
J. I. Penney,$^{1}$\thanks{E-mail: jip3@le.ac.uk}
 A. W. Blain,$^{1}$
D. Wylezalek,$^{2}$
N. A. Hatch,$^{3}$
C. Lonsdale,$^{4}$
\and A. Kimball,$^{5}$
R. J. Assef,$^{6}$
J.J. Condon,$^{4}$
P. R. M. Eisenhardt,$^{7}$
S. F. Jones,$^{8}$
\and M. Kim,$^{9,10}$
M. Lacy,$^{4}$
S. I. Muldrew,$^{1}$
S. Petty,$^{11}$
A. Sajina,$^{12}$
A. Silva,$^{12,13}$
\and D. Stern,$^{7}$
T. Diaz-Santos,$^{6}$
C-W. Tsai$^{14},$
J. Wu$^{14,15}$
\\
$^{1}$University of Leicester, Department of Physics and Astronomy, University Road, LE1 7RH, UK \\
$^{2}$Department of Physics and Astronomy, Johns Hopkins University, Baltimore, MD 21218, USA\\
$^{3}$School of Physics and Astronomy, University of Nottingham, University Park, NG7 2RD, UK\\
$^{4}$National Radio Astronomy Observatory, 520 Edgemont Road, Charlottesville, VA 22903, USA\\
$^{5}$National Radio Astronomy Observatory, 1003 Lopezville Road, Socorro, NM 87801, USA\\
$^{6}$N{\'u}cleo de Astronom{\'i}a de la Facultad de Ingenier{\'i}a y Ciencias, Universidad Diego Portales, Av. Ej{\'e}rcito Libertador 441, Santiago, Chile\\
$^{7}$Jet Propulsion Laboratory, California Institute of Technology, 4800 Oak Grove Drive, CA 91109, USA\\
$^{8}$Department of Space, Earth, and Environment, Chalmers University of Technology, Onsala Space Observatory, SE-43992, Onsala, Sweden\\
$^{9}$Korea Astronomy and Space Science Institute, Daejeon 305-348, Korea\\
$^{10}$Department of Astronomy and Atmospheric Sciences, Kyungpook National University, Daegu 702-701, Korea\\
$^{11}$NorthWest Research Associates, 4118 148th Ave NE, Redmond, WA 98052, USA\\
$^{12}$Department of Physics and Astronomy, Tufts University, 574 Boston Ave, MA 02155, USA\\
$^{13}$Observatory of Japan, National Institutes of Natural Sciences, 2-21-1 Osawa, Mitaka, Tokyo 181-8588, Japan\\
$^{14}$Department of Physics and Astronomy, University of California, Los Angeles, CA 90095, USA\\
$^{15}$National Astronomical Observatories, Chinese Academy of Sciences, 20A Datun Road, Chaoyang District, Beijing, 100012, China
}

\date{Accepted XXX. Received YYY; in original form ZZZ}

\pubyear{2018}

\begin{document}
\label{firstpage}
\pagerange{\pageref{firstpage}--\pageref{lastpage}}
\maketitle

\begin{abstract}
We have observed the environments of a population of 33 heavily dust obscured, ultra-luminous, high-redshift galaxies, selected using WISE and NVSS at $z>$1.3 with the Infra-Red Array Camera on the $Spitzer$ Space Telescope over $\rm5.12\,'\times5.12\,'$ fields. Colour selections are used to quantify any potential overdensities of companion galaxies in these fields. We find no significant excess of galaxies with the standard colour selection for IRAC colours of $\rm[3.6]-[4.5]>-0.1$ consistent with galaxies at $z>$1.3 across the whole fields with respect to wide-area $Spitzer$ comparison fields, but there is a $\rm>2\sigma$ statistical excess within $\rm0.25\,'$ of the central radio-WISE galaxy. Using a colour selection of $\rm[3.6]-[4.5]>0.4$, 0.5 magnitudes redder than the standard method of selecting galaxies at $z>$1.3, we find a significant overdensity, in which $\rm76\%$ ($\rm33\%$) of the 33 fields have a surface density greater than the $\rm3\sigma$ ($\rm5\sigma$) level. There is a statistical excess of these redder galaxies within $\rm0.5\,'$, rising to a central peak $\rm\sim2$--4 times the average density. This implies that these galaxies are statistically linked to the radio-WISE selected galaxy, indicating similar structures to those traced by red galaxies around radio-loud AGN.
\end{abstract}

\begin{keywords}
galaxies: evolution -- infrared: galaxies -- galaxies: active
\end{keywords}



\section{Introduction}
High-redshift galaxy protoclusters are the largest known cosmological structures, with 90$\%$ of their $\rm\sim10^{15} M_{\odot}$ mass expected within regions spanning $\rm\sim35\,$ $h^{-1} \,$Mpc \citep{YKChiang13,SIMuldrew15}. Simulations of protoclusters by \cite{YKChiang17} predicted significant excesses of star-forming galaxies in the central $\rm10-20\,Mpc$ of simulated protoclusters, suggesting these structures are important to understanding star formation in the early universe and the largest observed clusters in the local universe \citep{CCSteidel98, CCSteidel05}. However, given their extended nature it is observationally difficult to detect them \citep{SIMuldrew15}, with previous work using luminous Infra-Red (IR) galaxies to indicate potential sites of protoclusters to identify their properties \citep{OLeFevre96,PNBEST03,DW13,NHatch14}.

Some luminous IR galaxies harbour luminous Active Galactic Nuclei (AGN), at times placing them in the Ultra-luminous Infra-Red Galaxy (ULIRG\footnote{Here, LIRGs, ULIRGs and HyLIRGs are characterized by a total Infra-Red luminosity of $10^{11}$ $<$$L_{\rm8-1000\,\mu m}$ $<$$10^{12}$$\rm L_{\odot}$, $10^{12}$ $<$$L_{\rm8-1000\,\mu m}$ $<$$10^{13}$$\rm L_{\odot}$ and $\rm L_{8-1000\,\mu m}$$>$$10^{13}$$\rm L_{\odot}$ respectively}) regime \citep{DBSanders96,CJLonsdale06, CWTsai15}. A fraction of these ULIRGs have been identified using $Spitzer$ as ``Dust Obscured Galaxies" \cite[DOGs;][]{ADey08}, first uncovered owing to their high 24$\,\mu$m to R-band flux ratio ($\rm\frac{24\,\mu m}{R}>1000$) \citep{ADey08, DNarayanan10}. Hot Dust Obscured Galaxies \cite[Hot DOGs;][]{JWu12} are similar but rarer galaxies discovered using the Wide-Field Infra-Red Survey Explorer \cite[WISE;][]{ELWright10} All-Sky Survey, with red detections in the WISE W3 ($\rm12\,\mu m$) and W4 ($\rm22\,\mu m$) bands \citep{PRMEisenhardt12}. Hot DOGs exhibit higher dust temperatures ($\rm\gtrsim60K$) than DOGs \citep{JWu12, CRBridge13} and are powered by very massive, luminous AGNs \citep{JWu18}. Hot DOGs appear to reside within significantly overdense regions, whether traced by $\rm Ly\alpha$, IR-selected or sub-millimeter selected galaxies \citep{CRBridge13,SFJones14,RJAssef15}, implying they may be heating dust and spurring star formation. 

Sub-Millimeter Galaxies \cite[SMGs;][]{AWBlain02,ISmail02,LJTacconi08,CMCasey14} are a luminous, gas-rich population with high star formation rates that could be associated with earlier stages of Hot DOGs' evolution \citep{TRGreve05,AMSwinbank06,LJTacconi06}. Some DOGs have comparable rates of star formation and IR luminosities to SMGs \citep{DNarayanan10}. Further, Hot DOGs could be a link between SMGs and the optical quasar (QSO) population, seen during later evolutionary stages \citep{DBSanders88a,DBSanders88b,PFHopkins06}. Thus, there may be a link between these different populations related by their central AGN activity, star formation and environment \citep{DBSanders88a,DBSanders88b,CRBridge13,RJAssef15,ASilva15}.

Studies of powerful AGN at similar redshifts have revealed significant overdensities of near- and far-IR galaxies \citep{DStern03,JAStevens03,PNBEST03,CDeBreuck04,BPVenemans07,TRGreve07}. For example, investigations into the environment of the radio-loud Spiderweb galaxy \citep{GKMiley06}, one of the most massive and intensely studied galaxies \citep{JKurk00, LPentericci00}, suggest that the local density of SMGs is four times that of the field \citep{HDannerbauer14}. These data taken in the far-IR, including observations from Herschel's PACS and SPIRE and $Spitzer$'s Multiband Imaging Photometer \cite[MIPS;][]{MIPS}, suggest that the Spiderweb inhabits a region of intense star formation and potential mergers around a central, massive galaxy.

Systematic studies using IR measurements centred on radio-loud galaxies motivated by observations of the Spiderweb galaxy were made to understand the nature of these galaxies and their surroundings. The Clusters Around Radio-Loud AGN (CARLA) survey \citep{NHatch14,DW13,DW14} used $Spitzer$'s Infra-Red Array Camera \cite[IRAC;][]{GF04} observations on 420 fields containing Radio-Loud AGN (RLAGN), comparing the number of galaxies with specific colour selection criteria to fields in the $Spitzer$ UKIDSS Ultra-Deep Survey \cite[SpUDS;][]{SpUDS07} within several hundred kiloparsecs of the central galaxy \citep{DW13,DW14}. The survey found significant overdensities of IRAC colour-selected galaxies in the vicinity of these RLAGN, with $\rm\sim10\%$ of the $Spitzer$ fields containing densities $\rm>5\sigma$ with respect to comparison fields within a $\rm2.5\,'$ radius of the RLAGN.

Here we present $Spitzer$ IRAC observations of 33 fields centred on rare, ultra-luminous galaxies detected via the WISE All-Sky Survey using the W4 band for the 22$\rm\,\mu$m fluxes, and the NRAO $\rm1.4\,GHz$ VLA Sky Survey \citep[NVSS; ][]{JJCondon98} using more accurate radio positions from the higher-resolution FIRST survey where available \citep[see][for full details of the sample]{CJLonsdale15}. These galaxies have significant detections in the W3 and W4 bands (see Table~\ref{centrals}), and fainter detections in the W1 (3.5$\rm\,\mu m$) and W2 (4.6$\rm\,\mu m$) bands. Further selections are made for radio-loudness of $\rm-1< \log(F_{22\,\mu m}/F_{22\,cm})<\,0.1$ using NVSS and WISE. The radio-WISE galaxies are selected as radio-intermediate in power with compact emission to reduce the effect of synchrotron emission on IR-sub-millimeter fluxes and it is likely that these galaxies harbour radio jets from the core regions \citep{CJLonsdale16}. The sample could represent the early stages of luminous quasars, which include compact, young radio emission and powerful IR emission, consistent with dust enshrouded AGN, see \cite{CJLonsdale15}. Note that these galaxies are not as radio-loud as the CARLA targets but are typically more radio-loud than Hot DOGs, \citep{RJAssef15,CWTsai15}.

The SEDs of 49 of the radio-WISE selected galaxies using the Atacama Large Millimeter/Sub-Millimeter Array (ALMA; \cite{ALMA09}) \citep{CJLonsdale15} found 26 had detections in the $\rm870\,\mu$m band \citep{CJLonsdale15} and place them in the ULIRG and HyLIRG regimes, similar to Hot DOGs \citep{PRMEisenhardt12,JWu12,CRBridge13,SFJones14}, with dust temperatures $\rm\gtrsim30\,K$. Throughout this work, we refer to the targets as radio-WISE galaxies. The subset in this work, shown in Table~\ref{centrals}, were selected from the main catalogue \citep{CJLonsdale15} for spectroscopic redshifts $z>$1.3, with a maximum redshift of $z$=2.72. Radio-WISE galaxies have a significant overdensity of SMGs found at longer 850$\,\mu$m wavelengths in the surrounding $\rm\sim$$\rm1.5\,'$ radius \citep{SFJones15} with overdensities of 4--6 times that of the field. These overdensities were greater than that observed for fields containing Hot DOGs \citep{SFJones14}, with no evidence of angular clustering on scales out to $\rm1.5\,'$ from the target, and are expected to be radio-loud counterparts to Hot DOGs \citep{CJLonsdale15}. Furthermore, ALMA found overdensities of SMGs 10 times that of blank fields for 17 of the 49 galaxies within scales of $\rm\sim150\,$kpc ($\rm17\,''$) \citep{ASilva15}. The findings suggests that many of these HyLIRGs could inhabit dense unvirialised regions, given the levels of overdensity extending over large scales. Findings around these galaxies are consistent with simulations of large scale structure by \cite{YKChiang13} and \cite{SIMuldrew15}, which predict pre-virialised objects can be signposted by significant overdensities of dusty, star-forming galaxies.

It is of interest to understand how these extremely luminous and powerful radio-WISE selected galaxies interact with and affect their surrounding environment. IRAC colours are used to select galaxies at consistent redshifts and the surface densities of galaxies in the environment of radio-WISE galaxies. We characterise the nature of these fields using IRAC to greater than 10 times the depth of WISE observations. Section~\ref{data} describes the sample and the methods used to reduce the data and Section~\ref{results} discusses the findings. Throughout, we assume a cosmology of $H_{0}$ = 70 $\rm km$ $\rm s^{-1}$ $\rm\,Mpc^{-1}$, $\rm \Omega_{m}$ = 0.26 and $\rm \Omega_{\Lambda}$ = 0.74. All colours and magnitudes are displayed in AB-magnitudes, and IRAC1 and IRAC2 magnitudes are expressed as [3.6] and [4.5] respectively.

\section{$Spitzer$ Observations}\label{data}
\subsection{Data Reduction}\label{SExtractor}
We targeted 33 specific fields containing radio-WISE galaxies with redshifts $z>\rm1.3$ using $Spitzer$'s IRAC camera (PI: Carol Lonsdale, PID: 11013). Given the rarity of these galaxies, with sky-densities of 0.025$\,$deg$\rm^{-2}$, these galaxies do not lie within existing publicly available deep fields. Given their red colours and obscured nature, warm $Spitzer$ observations are the most sensitive to galaxies in the surrounding environment. Our observations follow a dithered pattern centred on the radio-WISE galaxy. Fourteen $\rm5.2\,'\times5.2\,'$ frames per field are imaged for $\rm\sim100\,s$ each and mosaicked to produce a deeper image of the field. This produces a square $\sim$30.8 arcmin$^{2}$ ($\rm5.55\,'\times5.55\,'$) two band image centred on the radio-WISE selected galaxy with a Point Spread Function (PSF) of the images of $\rm\sim1.95\,''$ for IRAC1 and $\rm\sim2.02\,''$ for IRAC2, found in the IRAC Instrument Handbook. The central $\rm5.12\,'\times5.12\,'$ region of the field was measured for $\rm\sim700\,$s for IRAC1 and $\rm\sim1000\,$s for IRAC2. 

The frames for each field were stacked and reduced using the MOPEX package \citep{MOPEX}, specifically designed for manipulating $Spitzer$ images. The pixel scale was set to $\rm0.6\,''$ in MOPEX and we used optimized parameters for deep IRAC images \citep{MLacy05}. A drizzling factor of 1.2 was added in the image interpolation stage to reduce cosmic-ray and bad-pixel effects. The first two frames for each field were also removed from the image mosaicking process due to higher levels of noise, reducing the time imaged for each radio-WISE field to $\sim500\,$s for IRAC1.

Source extraction was performed using Source Extractor \citep[henceforth SExtractor:][]{EBertin96} in single-image mode. Optimized SExtractor parameters for IRAC were taken from \cite{MLacy05} with a flux aperture diameter of $\rm4\,''$ ($\rm\sim2$ times the PSF of $Spitzer$). We convert from the native MJy sr$\rm^{-1}$ units of the image to $\rm \mu Jy$ $\rm pixel^{-1}$, using the conversion factor of $\rm8.4619\,\mu Jy$ $\rm pixel^{-1}$/MJy sr$\rm^{-1}$ for the $\rm0.6\,''$ pixel scale. We confirmed empirical aperture corrections by comparing the flux densities derived from SExtractor in the SpUDS field to those from the SpUDS catalogue \citep{SpUDS07}. These conversion factors were found to be 1.42 and 1.45 for [3.6] and [4.5] respectively, in agreement with \cite{DW13}. 

Finally, we determined the number of galaxies from the source extraction process that could be artifacts of either bright sources in the field or fluctuations in the background intensity. Bright sources can create diffraction effects, producing areas around the galaxy with greater levels of background intensity, which could create spurious galaxies within the fields. In general, we see $\rm\sim2$ objects per field which could be caused by these bright sources, representing $\rm<2\%$ of the galaxies in our fields. Galaxies at the edge of the field could be affected by the dithering pattern, which produces portions of the field imaged for less time. This could result in galaxies which may appear unrealistically blue/red in the catalogue. We therefore only analyse the uniformly covered central $\rm5.12\,'\times5.12\,'$ square ($\rm\sim7.15\,$Mpc$\rm^{2}$), centred on the radio-WISE galaxy to mask out these objects.

\begin{figure*} 
\captionsetup[subfigure]{labelformat=empty}
\subfloat[][]{\includegraphics[trim={0.7cm 0cm 0.9cm 1.2cm}, clip, width = 0.45\textwidth]{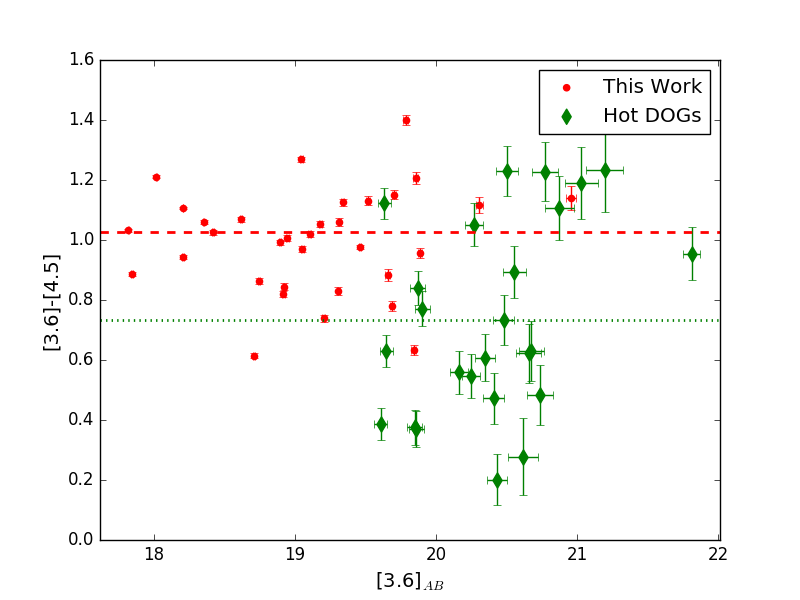}\label{compar1}}
\subfloat[][]{\includegraphics[trim={0.5cm 0cm 1.5cm 1.2cm},clip,width=0.45\textwidth]{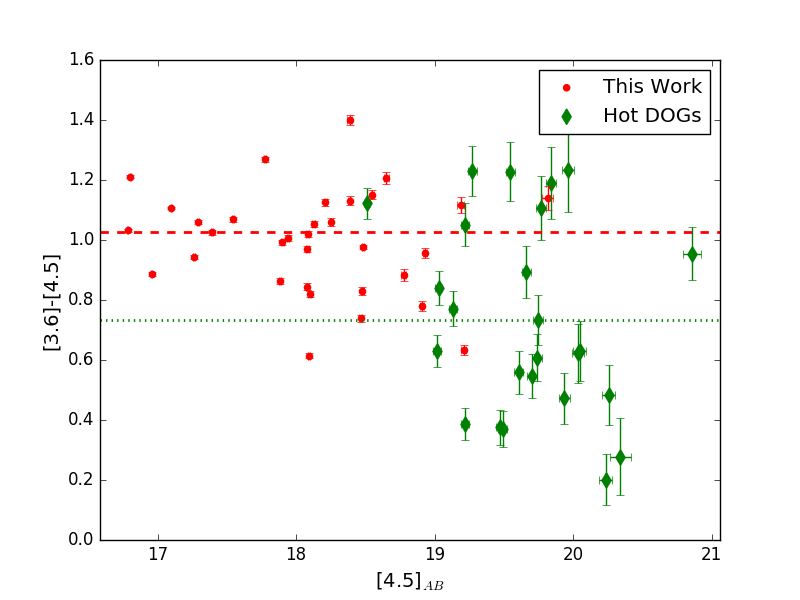}\label{compar2}}\\[-6ex]
\caption{Comparison of $Spitzer$ properties of the 33 central radio-WISE selected galaxies in this work, and 25 Hot DOGs \citep{JWu12} comparing the IRAC colour with [3.6] (left) and [4.5] (right). A red dashed and green dotted line represent the median $\rm[3.6]-[4.5]$ IRAC colours for the radio-WISE galaxies and Hot DOGs respectively. The radio-WISE selected galaxies in this work (red points), appear to have brighter IRAC magnitudes than the Hot DOGs (green diamonds). The radio-WISE galaxies have been imaged to a greater depth than these Hot DOGs, with $Spitzer$ observation times of $\rm\sim1000$s and $\rm\sim150$s respectively (see Section~\ref{SExtractor}).}
\label{comparfig}
\end{figure*}

\subsection{Properties of Radio-WISE Selected Galaxies}
We compare the $Spitzer$ properties of the central radio-WISE selected galaxies, listed in Table~\ref{centrals}, with those of Hot DOGs \citep{JWu12}, shown in Fig.~\ref{comparfig}. From Table~\ref{centrals}, we see that the IRAC [3.6] and [4.5] magnitudes of the radio-WISE selected galaxies are significantly brighter than the average magnitudes of red galaxies in the fields (<[3.6]$\rm_{AB}$>=21.34$\pm$0.97 and <[4.5]$\rm_{AB}$>=21.17$\pm$0.94): these galaxies are well measured in the rest-frame near-IR. They also possess redder IRAC $\rm[3.6]-[4.5]$ colours than average galaxies in the field, consistent with obscured galaxies. Comparing the IRAC $\rm[3.6]-[4.5]$ colours of the 33 radio-WISE galaxies in this work with the 25 Hot DOGs in \cite{JWu12}, we find slightly redder IRAC colours in the radio-WISE selected population using the same apertures and corrections, with a median colour of $\rm1.02\pm0.17$ for the radio-WISE galaxies, and $\rm0.73\pm0.33$ for the Hot DOGs. The radio-WISE selected galaxies are typically brighter and redder than the Hot DOG population. There is significant scatter in Fig.~\ref{comparfig} for the radio-WISE selected galaxies, with a Pearson correlation coefficient of 0.11 and $\rm-0.14$ for IRAC1 and IRAC2 respectively, suggesting that there is little correlation between their flux densities in the mid-IR and their $\rm[3.6]-[4.5]$ colour. There is slightly less scatter for the Hot DOG population, where the Pearson correlation coefficient is 0.38 and $\rm-0.24$ for IRAC1 and IRAC2 respectively, although there is no significant correlation between mid-IR flux density and $\rm[3.6]-[4.5]$ colour.

Using the WISE W3 ($\rm12\,\mu$m) and W4 ($\rm22\,\mu$m) data, we see that the median W3-W4 colour for the radio-WISE galaxies and Hot DOGs are $\rm1.20\pm0.34$ and $\rm1.79\pm0.27$ respectively. This suggests that the radio-WISE galaxies generally have bluer mid-IR colours than the Hot DOGs in \cite{JWu12}. Hot DOGs were also selected based on their $\rm W2-W3$ colour, and we see a redder median $\rm W2-W3$ colour for the Hot DOGs compared with the radio-WISE galaxies in this sample, with $\rm4.49\pm0.86$ and $\rm2.89\pm0.55$ for the Hot DOGs and radio-WISE galaxies respectively. There is a stronger correlation between the W3 and W4 bands and the W3-W4 colour for both classes of galaxy. For the radio-WISE galaxies, we find a correlation coefficient of 0.46 and -0.11 for W3 and W4 respectively. For the Hot DOGs in \cite{JWu12}, we find a correlation coefficient of 0.55 and -0.22 for W3 and W4 respectively. This suggests that, generally, the WISE colours are more correlated with their $\rm12\,\mu$m and $\rm22\,\mu$m flux densities than the IRAC flux densities for both galaxy classes.

Comparing the luminosity of the 30 radio-WISE galaxies in common with \cite{CJLonsdale15} with the luminosity estimates for the Hot DOGs in \cite{JWu12}, we find an average luminosity of $\rm2.7\times10^{13}\,L_{\odot}$ for the radio-WISE galaxies compared with $\rm6.1\times10^{13}\,L_{\odot}$ for the Hot DOGs. Comparing the correlation of the colour of the radio-WISE selected galaxies with the mid-IR luminosities listed in \cite{CJLonsdale15}, we find a correlation coefficient of 0.31 for $\rm[3.6]-[4.5]$ and 0.42 for W3-W4 colours, suggesting the redder IRAC and WISE colours are associated with brighter mid-IR luminosities. For the 9 Hot DOGs in \cite{JWu12} with sub-millimeter data, we see a correlation coefficient of 0.42 and -0.42 for IRAC and WISE colours respectively, suggesting that there is a correlation between the total luminosities of these galaxies and their mid-IR colours.

\begin{figure*} 
\captionsetup[subfigure]{labelformat=empty}
\subfloat[][]{\includegraphics[trim={0.7cm 0cm 0.9cm 0.3cm}, clip, width = 0.5\textwidth]{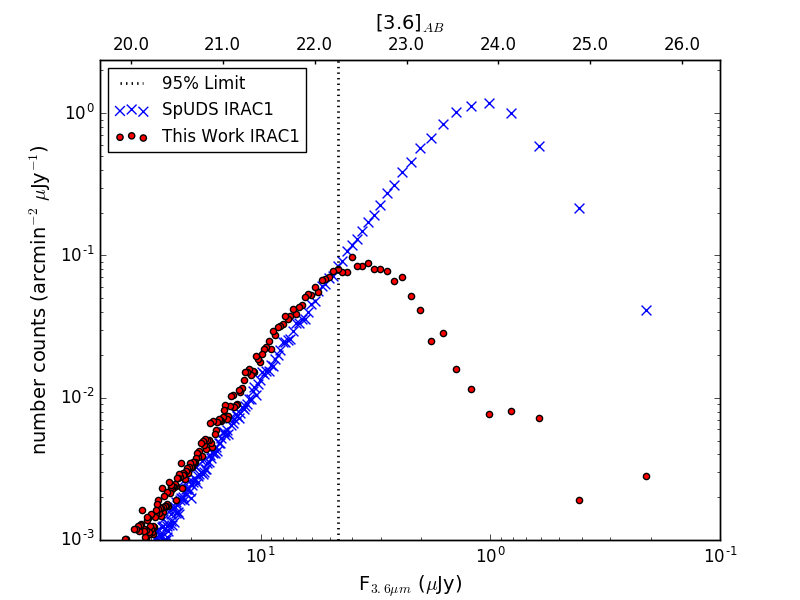}\label{comp1}}
\subfloat[][]{\includegraphics[trim={0.5cm 0cm 1.5cm 0.3cm},clip,width=0.5\textwidth]{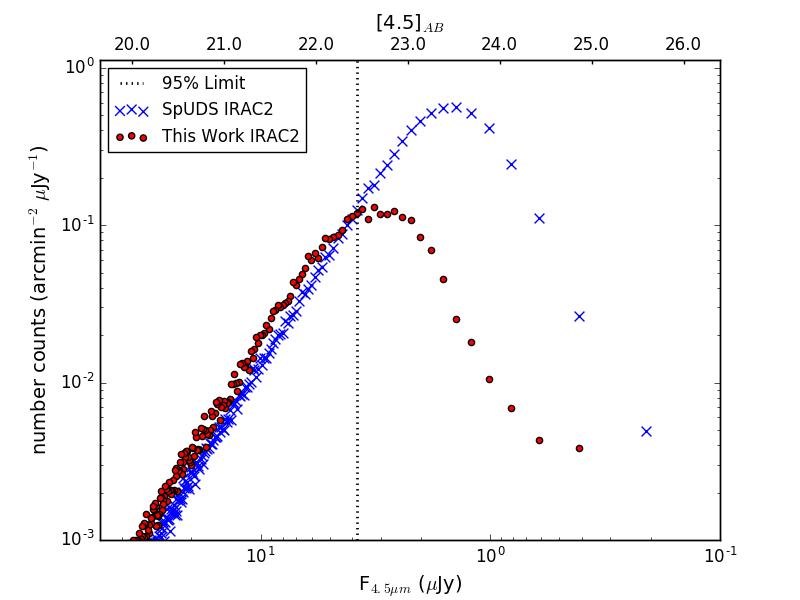}\label{comp2}}\\[-5ex]
\caption{Our completeness limits for IRAC1 (left) and IRAC2 (right). The blue crosses indicate the SpUDS data and the red points correspond to 33 stacked radio-WISE selected fields. A black dashed line has been included to indicate the $\rm95\%$ completeness limit of this work.}
\label{compfig}
\end{figure*}

\subsection{Completeness Limits}\label{completeness}
To determine the extent to which the survey could identify potential companion galaxies, we investigated the depth at which source detection was $\rm95\%$ complete. We analysed the number of galaxies detected in the radio-WISE selected fields, and compared them with the number of galaxies detected in SpUDS. Given that the SpUDS catalogue was created in a different manner to the source extraction method detailed above, the images of the SpUDS field was run through the same SExtractor method as our own fields for comparison.

From this, we determine the $\rm95\%$ completeness limits (see Fig.~\ref{compfig}) are [3.6]=22.23 and [4.5]=22.44, corresponding to source flux limits of 4.71$\rm\,\mu$Jy and 3.82$\rm\,\mu$Jy. The noise level was $\rm0.06\,\mu Jy$ ([3.6]=23.95) in IRAC1 and $\rm0.04\,\mu Jy$ ([4.5]=23.70) in IRAC2 in a $\rm4''$ diameter aperture.

\subsection{Object Selection Criteria}\label{selectioncrit}
Colour selection is used to isolate $Spitzer$ galaxies likely to be at $z>1.3$, matching the redshift range of the radio-WISE selected galaxies. \cite{Papovich08} demonstrated that, regardless of age or galaxy type, galaxies at $z>1.3$ can be selected by IRAC colour $\rm[3.6]-[4.5]>-0.1$ (or a flux density ratio of IRAC2 ($\rm F_{IRAC2}$) to IRAC1 ($\rm F_{IRAC1}$) of $\rm\frac{F_{IRAC2}}{F_{IRAC1}}>0.88$), based on models by \cite{GBruzual03}. This colour selection uses the $\rm1.6\,\mu m$ bump, a feature apparent in almost all galaxies, produced by the minimum opacity of the $\rm H^{-}$ ion in the atmospheres of cool stars \citep{TLJohn88}. Due to the relative placement of this bump and the IRAC channels, this colour selection method should be efficient at finding $z>$1.3 galaxies in the vicinity of radio-WISE selected galaxies \cite[see also][]{CSimpson99,DW13,RJAssef15}.

\begin{figure*}	
\captionsetup[subfigure]{labelformat=empty}
\subfloat[][]{\includegraphics[trim={0.5cm -0.85cm 0 -0.1cm},clip,height=6.3cm]{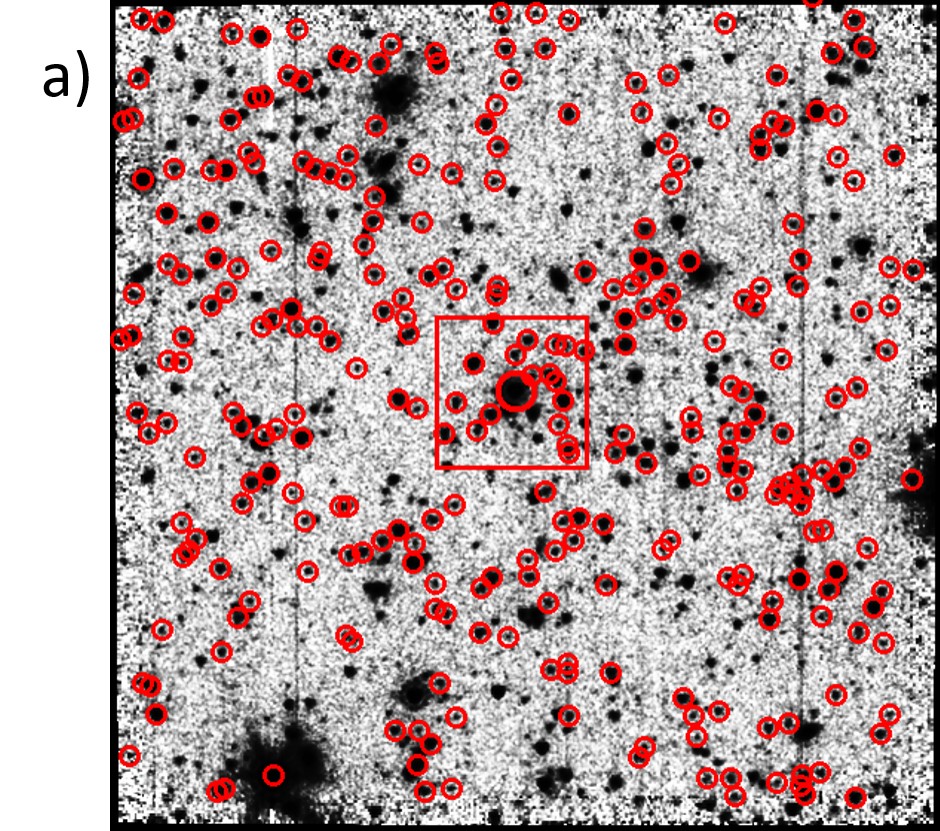}\label{pic_a}}
\subfloat[][]{\includegraphics[trim={-0.4cm 0.2cm 0.1cm -0.4cm},clip,height=6.6cm]{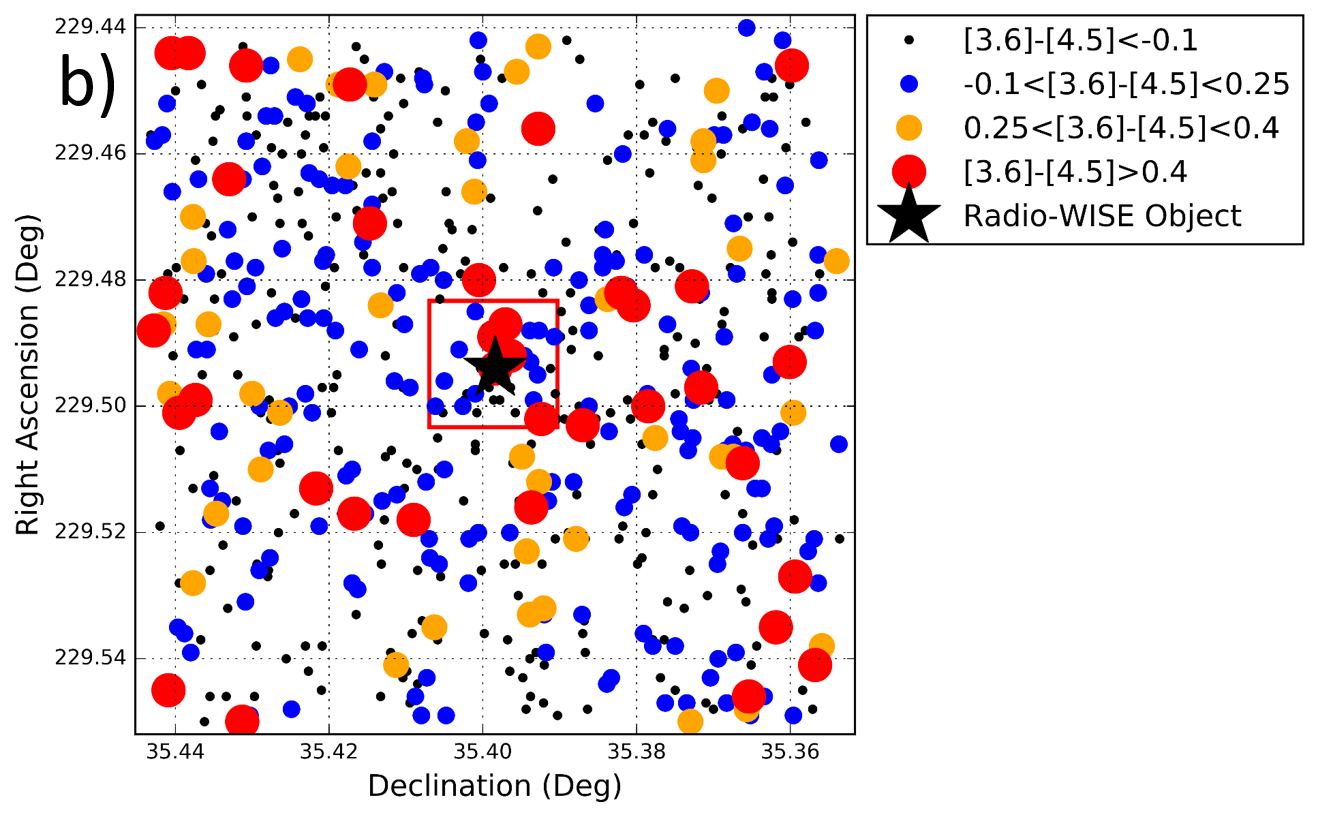}\label{map_b}}\\[-43ex]
\hspace*{12.8cm}\subfloat[][]{\includegraphics[height=2.1cm]{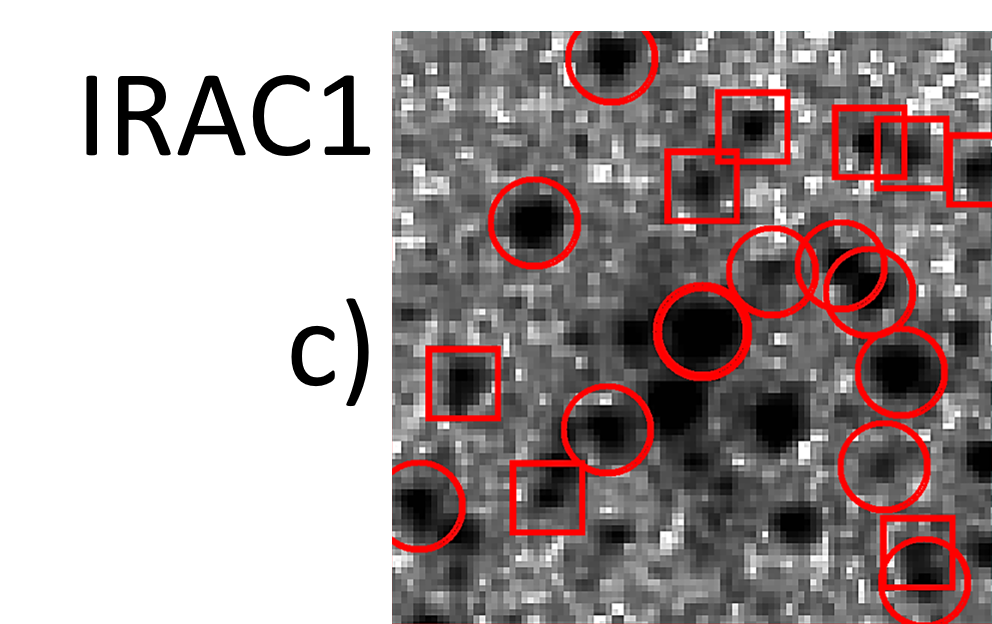}\label{inset_c}}\\[-7ex]
\hspace*{12.8cm}\subfloat[][]{\includegraphics[height=2.1cm]{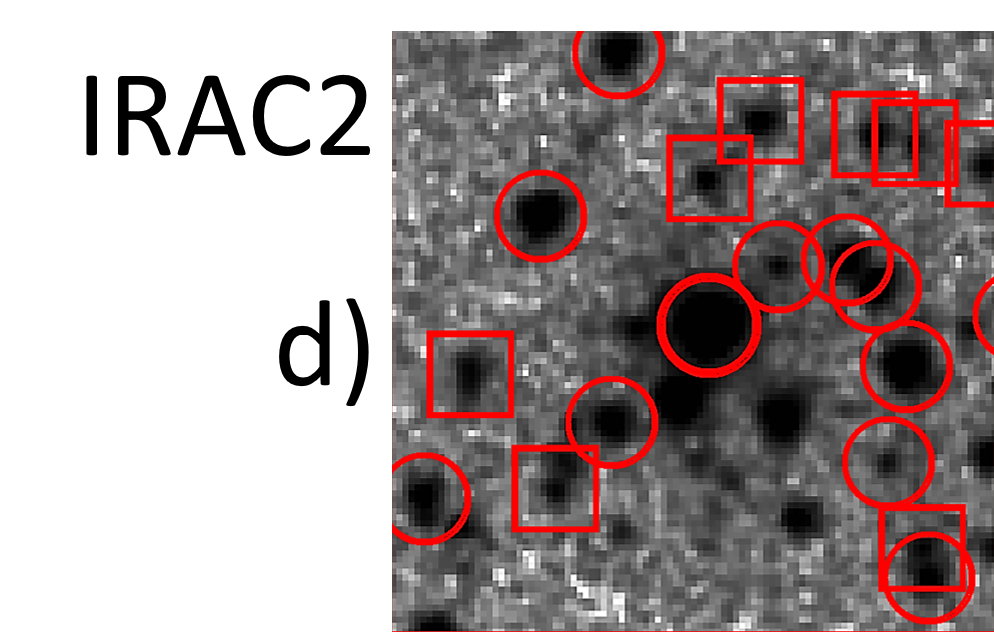}\label{inset_d}}\\[-4ex]
\caption{\protect\subref{pic_a} A typical IRAC2 $\rm5.55\,'\times5.55\,'$ field, centered on W1517 indicated by a large bold circle. Catalogued galaxies whose IRAC colours $\rm[3.6]-[4.5]>-0.1$, above the completeness limits discussed in Section~\ref{completeness} are marked by red circles. \protect\subref{map_b} Map of the same field with positions marked by the IRAC colours of the galaxies. \protect\subref{inset_c} Central $\rm1\,'\times1\,'$ ($\rm\sim500\,kpc\times500\,kpc$) region centred on the radio-WISE selected galaxy for IRAC1. \protect\subref{inset_d} Central $\rm1\,'\times1\,'$ region centred on the radio-WISE selected galaxy for IRAC2. For the inset regions, galaxies with IRAC1 magnitudes $\rm21<$[3.6]$\rm<22$ and IRAC colours $\rm[3.6]-[4.5]>-0.1$ are boxed, objects within the completeness limits and IRAC colours $\rm[3.6]-[4.5]>-0.1$ are circled and the central radio-WISE selected galaxy is shown by a large, bold ring. The inset regions are highlighted by the box in \protect\subref{pic_a} and \protect\subref{map_b}.}
\label{picfig}
\end{figure*}

Galaxies are selected above the completeness limits determined in Section~\ref{completeness} with these IRAC colours. Candidate galaxies with IRAC2 magnitudes within the completeness limits, but fainter for IRAC1, were also accepted within the colour limit $\rm[3.6]-[4.5]>-0.1$ to ensure a complete list of redder galaxies. Galaxies selected using $\rm[3.6]-[4.5]>-0.1$ (henceforth Group-I galaxies) appear to be distributed fairly uniformly across the $\rm5.12\,'\times5.12\,'$ field, with no obvious clustering around the central radio-WISE selected galaxy, as shown for a typical example in Fig.~\ref{picfig}\subref{pic_a}. Spectroscopic redshifts would be needed to reveal the three dimensional distribution of galaxies across the field, given the colour selection is consistent only with the detections having a redshift $z>$1.3. Galaxies in this catalogue will not all be at the same redshift as the radio-WISE selected galaxies. 

In Section~\ref{results} we compare the surface density, radial distribution and other features of these galaxies with large comparison fields. In the 33 radio-WISE selected fields, we detect $\rm\sim7400$ Group-I galaxies within the $\rm95\%$ completeness limits ($\rm\sim225$ galaxies per field).

\subsection{SpUDS and S-COSMOS Comparison Fields}\label{comparison}
SpUDS (PI: J. Dunlop) is a $Spitzer$ Cycle-4 legacy program, observing $\rm\sim1\,deg^{2}$ in the UKIDSS UDS field using IRAC and MIPS. SpUDS has a $\rm3\sigma$ depth of $\rm1\,\mu Jy$ \cite[$\rm m_{AB}=24$, ][]{KICaputi11}. S-COSMOS (PI: D. Sanders) is a $Spitzer$ Cycle-2 legacy program, covering the $\rm2\,deg^{2}$ of the Cosmic Evolution Survey (COSMOS) field, reaching to $\rm5\sigma$ depths of $\rm1\,\mu Jy$ \cite[$\rm m_{AB}$=24,][]{NScoville07}. These comparison fields are 3--5 times deeper than our survey and are complete at the $\rm95\%$ flux density limits for this work, making them ideal for comparison with densities of galaxies in the radio-WISE selected fields.

\begin{figure}        
\includegraphics[trim={0.4cm 0cm 1.5cm 1.2cm},clip,width=\columnwidth]{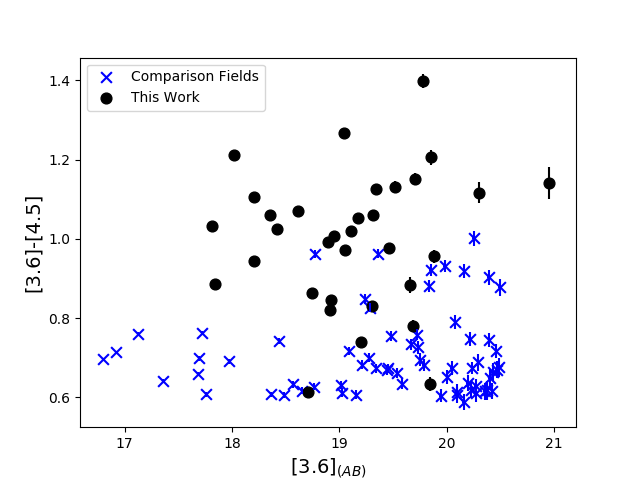}\\[-5ex]
\caption{Comparison of the $Spitzer$ IRAC properties between the radio-WISE selected galaxies and the central galaxies in the comparison fields (Section~\ref{comparison}). Radio-WISE selected galaxies are represented by black points, and centred galaxies from the SpUDS and S-COSMOS fields are represented by blue crosses.}
\label{centralsvscentralsfig}
\end{figure}

\begin{figure}        
\includegraphics[trim={0.8cm 0cm 1.5cm 1.2cm},clip,width=\columnwidth]{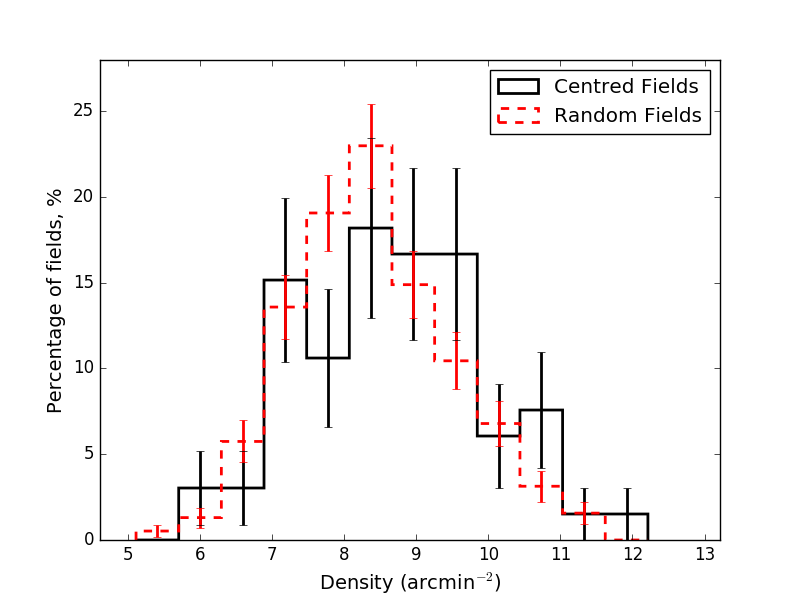}\\[-6ex]
\caption{Density of the comparison fields with color selection $\rm[3.6]-[4.5]>-0.1$. Fields centred randomly are shown in dashed red and those centred on galaxies with similar IRAC properties to the radio-WISE selected galaxies are shown in black.}
\label{papcentrevsrandomfig}
\end{figure}

To compare with the fields centred on radio-WISE galaxies we cut out fields centred on galaxies with IRAC colours $\rm[3.6]-[4.5]>0.58$, $\rm16<[3.6]<20.97$ and $\rm[4.5]<19.85$ in the SpUDS and S-COSMOS fields, as shown in Fig.~\ref{centralsvscentralsfig}. Fields were also selected to have the same $\rm5.12\,'\times5.12\,'$ ($\sim$26.2\,arcmin$\rm^{2}$) dimensions as the radio-WISE selected fields. This produced 15 and 51 fields in SpUDS and S-COSMOS respectively, which were combined in all investigations as we saw expected levels of variance between these blank fields. No radio-WISE selected galaxies are found in either of these comparison fields. As shown in Fig.~\ref{centralsvscentralsfig}, the median [3.6] magnitude and $\rm[3.6]-[4.5]$ IRAC colour for this control sample is 19.67 and 0.68 respectively, whereas the median values for the radio-WISE selected targets are 19.18 and 1.03 respectively. This suggests that the comparison sample is much bluer and fainter than the radio-WISE targets, showing that objects with similar IRAC properties to these rare, luminous radio-WISE selected galaxies are not observed in deep, wide-area, blank fields.

Given the lack of comparison fields unlike the non-overlapping, randomly-placed SpUDS fields in \cite{DW13}, we compared the density of Group-I galaxies in our fields to randomly-placed, independent pointings in the SpUDS and S-COSMOS fields of the same $\rm5.12\,'\times5.12\,'$ area, which produced 96 and 288 independent $\rm5.12\,'\times5.12\,'$ fields respectively. Fig.~\ref{papcentrevsrandomfig} compares the difference in density between the fields centred on galaxies with similar colours to the radio-WISE galaxies and to randomly placed fields. Using a Kolmogorov-Smirnov test (K--S test) we find a K--S statistic of 0.17 with a $p$-value of 0.06 for the two subsets of the comparison fields. A result with a $p$-value $\rm>0.05$ suggests that we cannot rule out that the two samples are drawn from the same underlying distribution and thus do not possess a significant excess. There appears to be little difference when using the fields centred on galaxies with similar IRAC properties to the radio-WISE galaxies to randomly selected fields, although we cannot rule out the possibility that these subsets are drawn from a different underlying distribution. We choose to use the more numerous randomly-centred fields when comparing to the radio-WISE selected fields to increase the number of comparison fields.

\begin{figure*} 
\captionsetup[subfigure]{labelformat=empty}
\subfloat[][]{\includegraphics[trim={0.2cm 0cm 0.2cm 0.2cm}, clip, width = 0.5\textwidth]{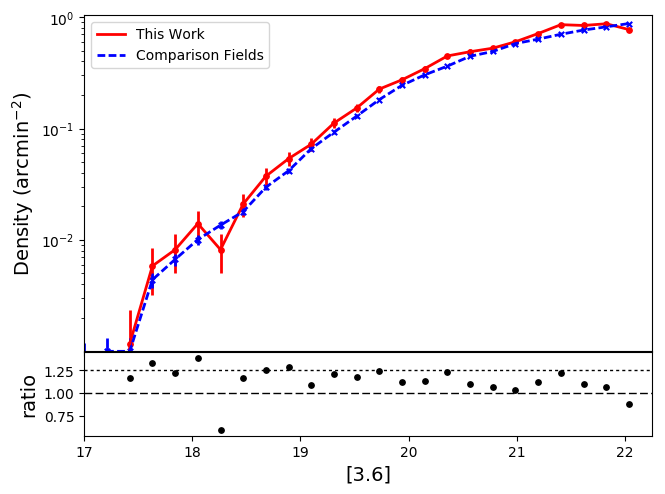}\label{Mag1}}
\subfloat[][]{\includegraphics[trim={0.2cm 0cm 0.2cm 0.2cm},clip,width=0.5\textwidth]{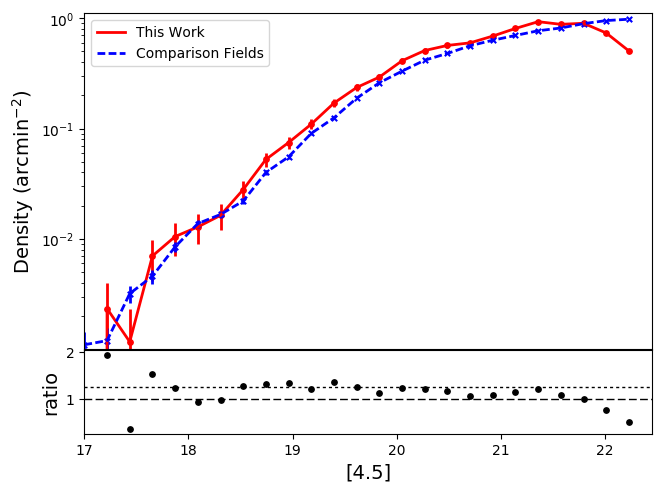}\label{Mag2}}\\[-5ex]
\caption{Comparison of the density of IRAC-selected galaxies in the 33 $\rm5.12\,'\times5.12\,'$ radio-WISE selected fields (solid red) in arcmin$\rm^{-2}$ mag$^{-1}$ with respect to their magnitudes in [3.6] (left) and in [4.5] (right), with comparison fields indicated in dashed blue. Below each figure is a density ratio of the radio-WISE fields against the comparison fields. A dashed and dotted line denote a $\rm1\times$ and $\rm1.25\times$ overdensity.}
\label{magnitudefig}
\end{figure*}

\section{Results}\label{results}
\subsection{Red-Selected $Spitzer$ Galaxies}\label{papovichobj}
The relative density of Group-I IRAC-selected galaxies in $\rm5.12\,'\times5.12\,'$ fields in the radio-WISE, SpUDS and S-COSMOS fields as a function of IRAC colour $\rm[3.6]-[4.5]>-0.1$ is shown in Fig.~\ref{magnitudefig}. 

We see a modest overdensity of $\rm\sim10\%$ across the magnitude range within the completeness limits of this work in both bands in the radio-WISE selected fields. Removing the radio-WISE galaxies we see the largest excess ($\rm\sim25\%$) in the magnitude range of $\rm18.5<$ m$_{AB}<20.5$ for both [3.6] and [4.5]. Cosmic variance is a significant factor in the comparison fields, expected to scatter the density of galaxies found in $\rm5.12\,'\times5.12\,'$ fields by $\rm\sim30\%$ \citep{MTrenti08}, greater than the observed excess in the radio-WISE selected fields. Given that the radio-WISE selected fields have been averaged, this suggests that the observed overdensity is unlikely to be caused by cosmic variance; these radio-WISE selected fields are modestly overdense. We see a reduced density of galaxies at fainter magnitudes in IRAC2 ([4.5]$\rm\gtrsim21.7$), likely due to the completeness limits (see Fig.~\ref{compfig}).

The excess of $\rm[3.6]-[4.5]>-0.1$ galaxies is much less than found using the same IRAC colour selection around radio-loud galaxies \citep{DW13}, suggesting that the environments of the radio-WISE galaxies are less significantly overdense in the IRAC bands than galaxies with bright radio emission. It is unlikely that this lack of excess density is due to the depth of observations in this work, given the relatively uniform overdensity across most magnitudes of this study, as shown in Fig.~\ref{magnitudefig}. The lack of a large excess relative to the comparison fields (10-$\rm25\%$ in comparison to the $\rm\sim4$-$\rm6\times$ seen in \cite{SFJones15} for sub-millimeter galaxies) applies across a wide magnitude range in this study, suggesting there is no specific magnitude range associated with the radio-WISE target galaxy.

\begin{figure}        
\includegraphics[trim={0.8cm 0cm 1.5cm 1.2cm},clip,width=\columnwidth]{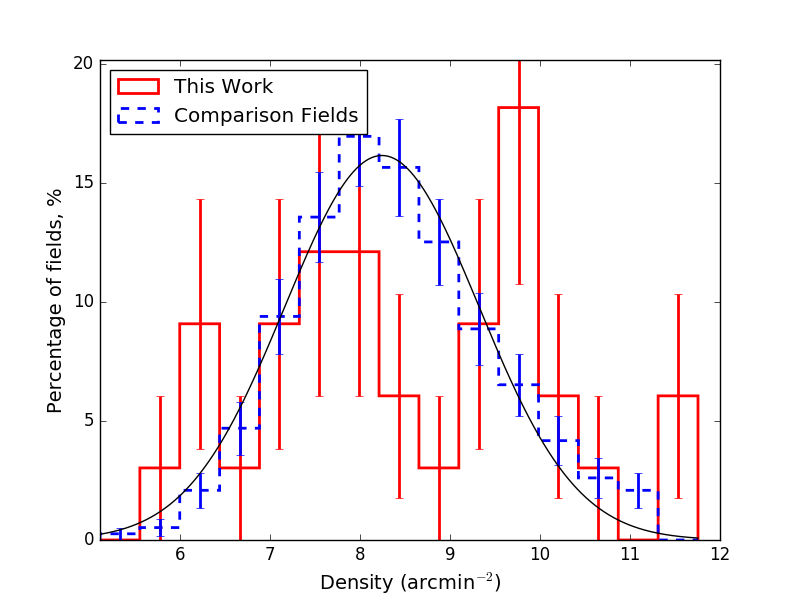}\\[-6ex]
\caption{Density of galaxies across the 33 $\rm5.12\,'\times5.12\,'$ fields compared to 383 comparison fields with the same dimensions. Radio-WISE selected fields are indicated in solid/red and comparison fields in dashed/blue. A Gaussian distribution has been fitted to the comparison fields in black.}
\label{papnumfield}
\end{figure}

To investigate the origin of the modest $\rm10\%$ overdensity compared to previous work on active galaxies, we determine the density field by field. We compare the distribution of galaxies per field to the number of similar galaxies in the comparison fields, in Fig.~\ref{papnumfield}. A Gaussian distribution was fitted to the comparison fields, giving 8.31$\pm$1.10 arcmin$\rm^{-2}$. $\rm55\%$ of the radio-WISE selected fields exceed the mean density and $\rm36\%$ are denser than the mean plus $\rm1\sigma$ density of the SpUDS fields. In comparison, \cite{DW13} found a more significant overdensity across a sample of 420 fields centred on RLAGN, with $\rm92\%$ of fields denser than the comparison sample mean, significantly greater than the excess around the radio-WISE galaxies. Using a K--S test and comparing the results of this investigation to the comparison fields, we find a K--S statistic of 0.22 with a $p$-value of 0.08. The radio-WISE selected fields are not significantly denser in IRAC-selected galaxies than the comparison fields.

\begin{figure}  
\includegraphics[trim={0.8cm 0cm 1.5cm 1.2cm},clip,width=\columnwidth]{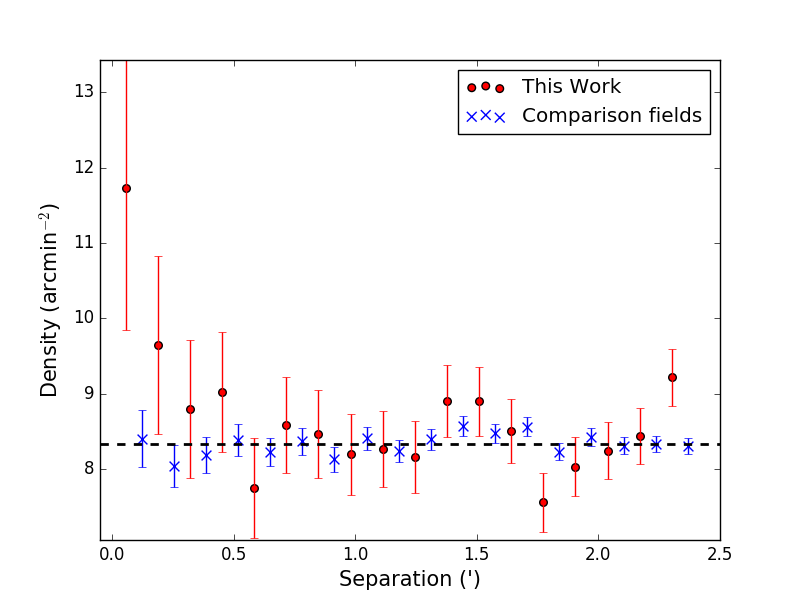}\\[-5ex]
\caption{The radial distribution of IRAC-selected galaxies away from the radio-WISE selected galaxy using the entire magnitude range of this work ([4.5]$\rm<22.45$). Galaxies have IRAC colour $\rm[3.6]-[4.5]>-0.1$, out to a $\rm2.5\,'$ radius from the centre of the field. Radio-WISE selected fields are represented by red circles and the comparison fields are indicated by blue crosses. The dashed line denotes the average density of the comparison fields for reference.}
\label{papradialfig}
\end{figure}

We investigate the radial distribution of these Group-I galaxies with respect to the central radio-WISE selected galaxies to determine if there is any spatial correlation between these galaxies and the central galaxy. All 33 fields were combined to provide a distribution of Group-I galaxies away from the central radio-WISE galaxy position, which was excluded from this investigation. The same distribution was constructed around the random positions for the comparison fields. As shown in Fig.~\ref{papradialfig}, there is no overdensity beyond $\rm\sim0.25\,'$ from the central galaxy. Towards the centre of the field, however, we see a rise in the number of Group-I galaxies $\rm\sim1.33\pm0.20$ times the level of the comparison fields. This implies possible clustering between these Group-I galaxies and the central galaxy on small scales of $\rm\lesssim0.25\,'$. This overdensity towards the centre of the field could be linked to the $\rm\sim10\%$ overdensity seen in Fig.~\ref{magnitudefig}. It should be noted that the result in Fig.~\ref{papradialfig} is obtained statistically over all 33 fields and does not represent the radial distribution of Group-I galaxies in each individual field. The peak in the figure corresponds to an average excess of $\rm\sim2$--3 galaxies per field within $\rm0.25\,'$ of the radio-WISE galaxy.

\begin{figure*} 
\captionsetup[subfigure]{labelformat=empty}
\subfloat[][]{\includegraphics[trim={0.7cm 0cm 0.9cm 1.2cm}, clip, width = 0.33\textwidth]{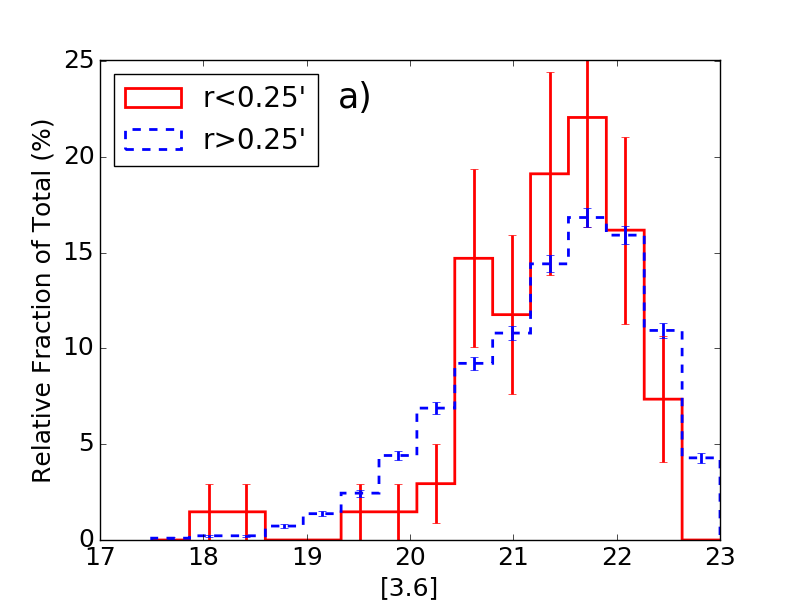}\label{innerIRAC1fig}}
\subfloat[][]{\includegraphics[trim={0.5cm 0cm 1.5cm 1.2cm},clip,width=0.33\textwidth]{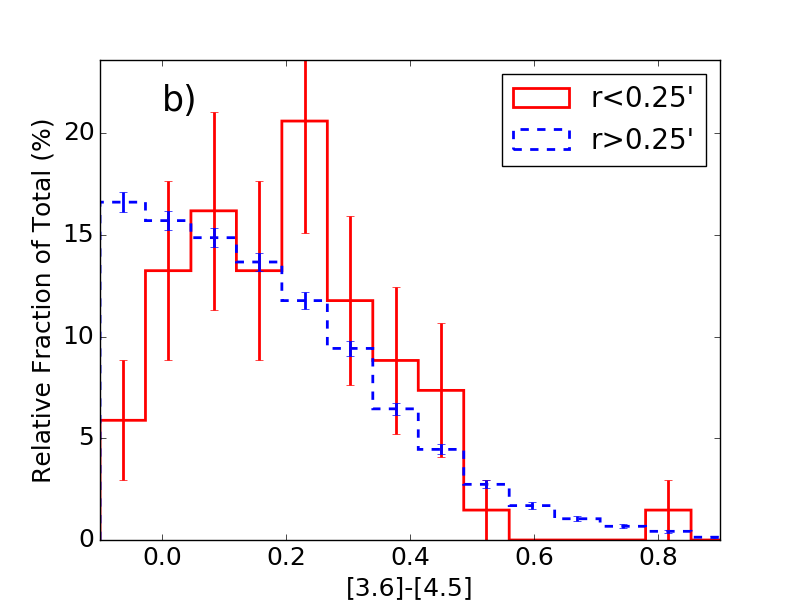}\label{innercolorfig}}
\subfloat[][]{\includegraphics[trim={0.2cm 0cm 1.5cm 1.2cm},clip,width=0.33\textwidth]{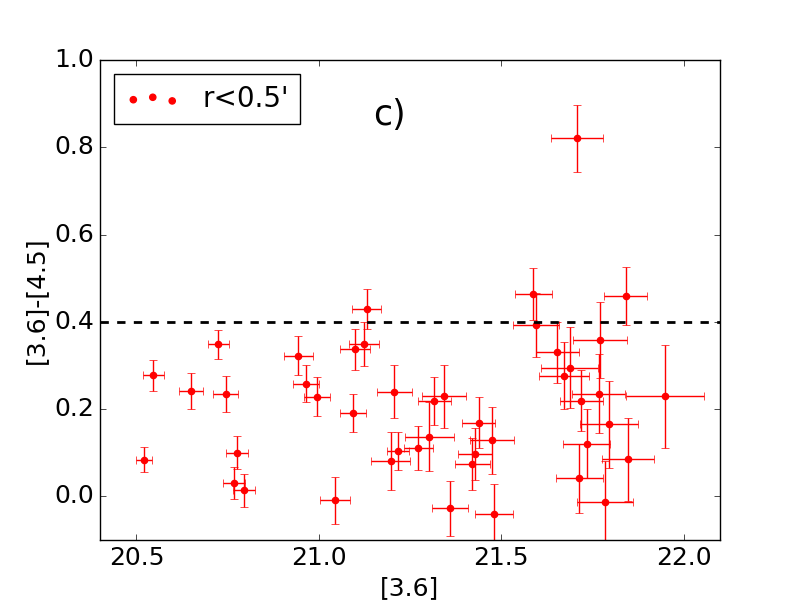}\label{innercolor_vs_mag1}}\\[-4ex]
\caption{Comparison of the properties of all IRAC selected galaxies summed across all 33 radio-WISE selected fields within $\rm0.25\,'$ of the central radio-WISE galaxy to the rest of the field: \protect\subref{innerIRAC1fig} compares the relative fraction of galaxies in the inner $\rm0.25\,'$ radius from the radio-WISE galaxy (red line) to the relative fraction of galaxies at radii $\rm>0.25\,'$ (blue dashed) as a function of [3.6]. \protect\subref{innercolorfig} compares the $\rm[3.6]-[4.5]$ colours of the galaxies in the peak to the [3.6] (red line) to the galaxies beyond $\rm0.25\,'$ separation from the radio-WISE galaxy (blue dashed). \protect\subref{innercolor_vs_mag1} shows the colors of galaxies in the inner $\rm0.25\,'$ radius from the radio-WISE galaxy against their colour. A black line has been added to denote the minimum colour of the radio-WISE galaxies.}
\label{innerfigs}
\end{figure*}

To determine whether this peak is produced by galaxies in a specific range of magnitudes or IRAC colours, we show the distribution of the [3.6] and the $\rm[3.6]-[4.5]$ colours of the galaxies in the inner $\rm0.25\,'$ in Fig.~\ref{innerIRAC1fig}. The figure shows that there is a potential modest excess of galaxies within the magnitude range of $\rm20.5<$[3.6]$\rm<22$. This excess is not significantly greater than the density for galaxies at separations $\rm>0.25\,'$ from the radio-WISE galaxies for most values of [3.6]. Comparing the two subsets, we find a K--S statistic of 0.12 with a $p$-value of 0.29 suggesting that these subsets are likely drawn from the same distribution. Comparing the IRAC colours (see Fig.~\ref{innercolorfig}), we see no significant excess of galaxies within any colour range except at [3.6]-[4.5]$\rm\sim0.2$. The distribution of these subsets are different, the density of galaxies in each colour bin rising between $\rm-0.1<[3.6]-[4.5]<0.2$ within $\rm0.25\,'$, in contrast to the rest of the field which shows a reduction in density with redder IRAC colour. Using a K--S test for these subsets, we find a K--S statistic of 0.17 with a $p$-value of 0.03, suggesting that we cannot accept the null hypothesis that these samples are drawn from the same distribution. From Fig.~\ref{innercolor_vs_mag1}, we find only $\rm\sim10\%$ of the galaxies in the inner $\rm0.25\,'$ exhibit $\rm[3.6]-[4.5]>0.4$, in a similar colour range to the radio-WISE galaxies, suggesting that the peak in Fig.~\ref{papradialfig} is not composed of galaxies with similar IRAC colours to the radio-WISE galaxies. These results show that the excess of galaxies in the local $\rm0.25\,'$ environment of the radio-WISE galaxies are faint but not exceptionally red.

\begin{figure}        
\includegraphics[trim={0.5cm 0cm 1.5cm 1.2cm},clip,width=\columnwidth]{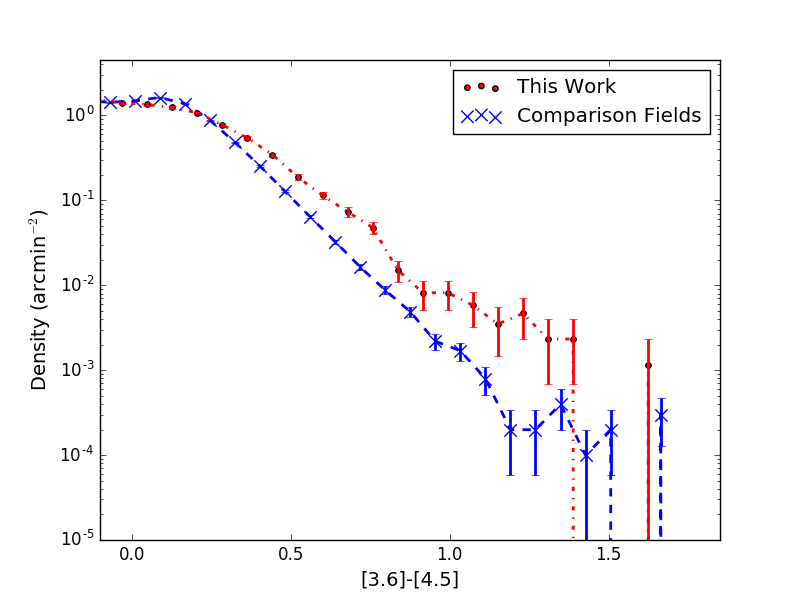}\\[-6ex]
\caption{Galaxies arcmin$^{-2}$ as a function of IRAC ($\rm[3.6]-[4.5]$) colours. Red circular points indicate radio-WISE selected fields, excluding the central radio-WISE galaxy, blue crosses indicate the comparison fields.}
\label{colorfig}
\end{figure}

\subsection{Redder-Selected $Spitzer$ Galaxies}\label{redderobjs}
To investigate whether the modest overdensity shown in Fig.~\ref{magnitudefig} can be linked to bluer or redder IRAC-selected galaxies, we determined the density of galaxies with respect to their $\rm[3.6]-[4.5]$ colour, excluding the central radio-WISE selected galaxies. A distribution of the density of red galaxies is shown in Fig.~\ref{colorfig}. The density of redder galaxies drops off more steeply at $\rm[3.6]-[4.5]>0.4$ for the SpUDS and S-COSMOS fields than in the radio-WISE fields. This is similar to the findings from CARLA \citep{EACooke16}, that there was an increasing fraction of redder galaxies around their target than in blank fields. The radio-WISE selected fields contain a greater density of increasingly red IRAC galaxies in their fields. This suggests that these radio-WISE selected galaxies reside in fields containing an excess of redder IRAC galaxies on $\rm5.12\,'\times5.12\,'$ scales.

\begin{figure}        
\includegraphics[trim={0.5cm 0cm 1.5cm 1.2cm},clip,width=\columnwidth]{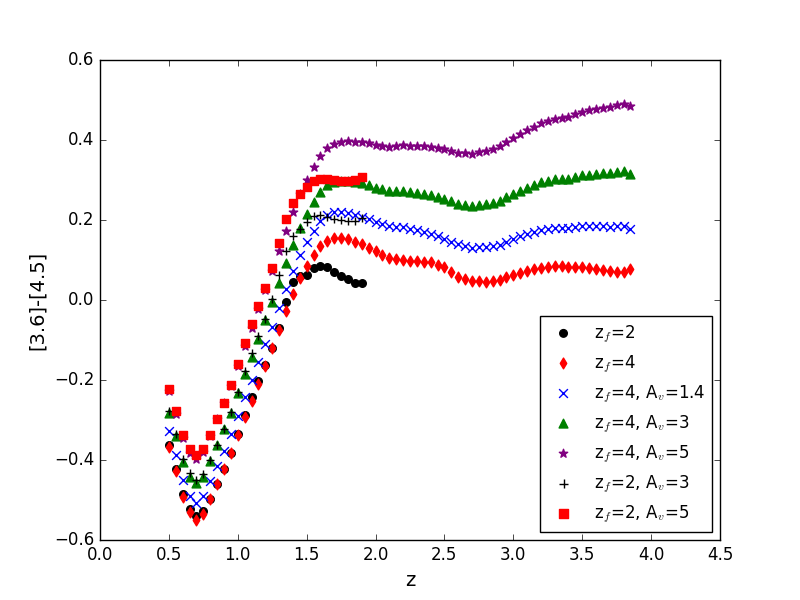}\\[-6ex]
\caption{Illustration of the colour dependence with redshift for different galaxy models at different formation redshifts, $z_{f}$. We use stellar libraries to model these galaxies \protect\citep{GBruzual03}, assuming a Salpeter initial mass function and a single exponentially decaying burst of star formation at $\tau$=$\rm0.1\,$Gyr. The models were generated using $EzGal$ \citep{CLMancone12} with additional dust extinction using a Calzetti law \protect\citep{DCalzetti00} for values of $\rm A_{v}$ to model the $\rm[3.6]-[4.5]$ colour for different extinction levels.}
\label{ezgalfig}
\end{figure}

Using models of galaxy SEDs, we attempt to probe the nature of the redder IRAC colours associated with these galaxies. For galaxies consistent with the same redshift as the targets, the excess galaxies are expected to be even redder than the models for extinctions as high as $\rm A_{v}=5$. Given a general extinction for galaxies of $\rm A_{v}\sim1.1$ \citep{PSklias14}, these redder-IRAC galaxies are significantly more dust obscured than typical galaxies. As Fig.~\ref{ezgalfig} shows, these colours would be consistent across the redshift range of the targets, such that the redder-IRAC colours are unlikely to be caused by redshift alone.

These redder-selected galaxies (henceforth Group-II galaxies) have similar IRAC $\rm[3.6]-[4.5]$ colours to the radio-WISE selected galaxies, indicating greater hot dust emission or higher levels of obscuration than field galaxies at $z>$1.3. We derive the same set of results as Section~\ref{papovichobj} for these Group-II galaxies to investigate their overdensity, using a colour selection of $\rm[3.6]-[4.5]>0.4$ (or a flux density ratio of $\rm\frac{F_{IRAC2}}{F_{IRAC1}}>\frac{7}{5}$). This colour cut was chosen to include the range of IRAC colours where there is an excess density of redder galaxies in the radio-WISE fields, highlighted in Fig.~\ref{colorfig}. It should be noted that Group II galaxies are a subset of Group I galaxies, and as shown in Fig.~\ref{innercolor_vs_mag1}, the peak in Fig.~\ref{papradialfig} is not composed of the Group-II galaxies.

\begin{figure}        
\includegraphics[trim={0.5cm 0cm 1.5cm 1.2cm},clip,width=\columnwidth]{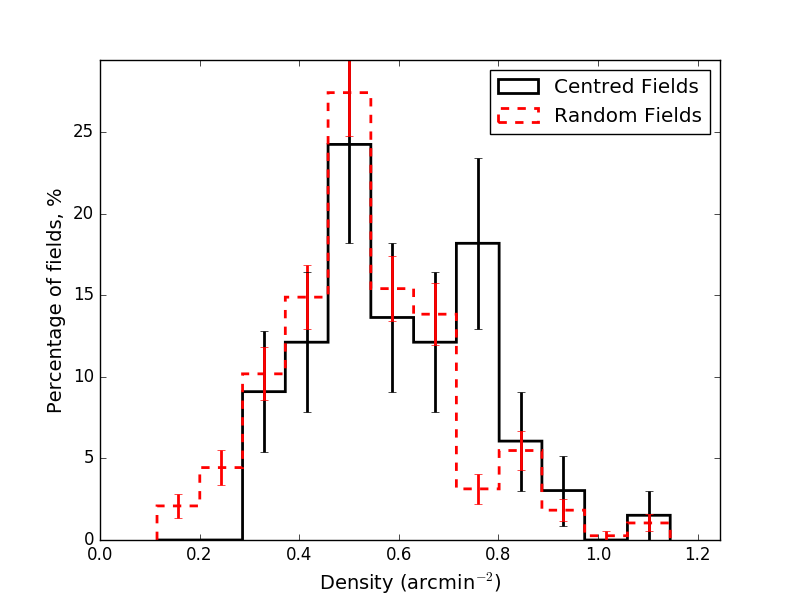}\\[-6ex]
\caption{Comparison of the density of SpUDS and S-COSMOS fields with color selection $\rm[3.6]-[4.5]>0.4$ between fields centred randomly and those centred on galaxies with similar IRAC properties to the radio-WISE selected galaxies. Randomly centred fields are shown in dashed red, while fields centred on galaxies are in black.}
\label{redcentrevsrandomfig}
\end{figure}

First, we determined whether the distribution of Group-II galaxies was different for fields centred on galaxies with similar IRAC properties to the radio-WISE galaxies compared with randomly-centred fields in the SpUDS and S-COSMOS fields. We investigated the density field to field, as shown in Fig.~\ref{redcentrevsrandomfig}, after the central galaxies in the fields were removed. There does not appear to be a significant difference in the density of these Group-II galaxies between the two comparison fields. A K--S test gives a K--S statistic of 0.17 with a $p$-value of 0.07, suggesting that these subsets are drawn from the same underlying distribution. To increase the number of comparison fields, we will use the randomly-centred SpUDS and S-COSMOS fields to compare with the radio-WISE selected fields for the Group-II galaxies.

\begin{figure} 
\includegraphics[trim={1cm 0cm 1.5cm 1.2cm},clip,width=\columnwidth]{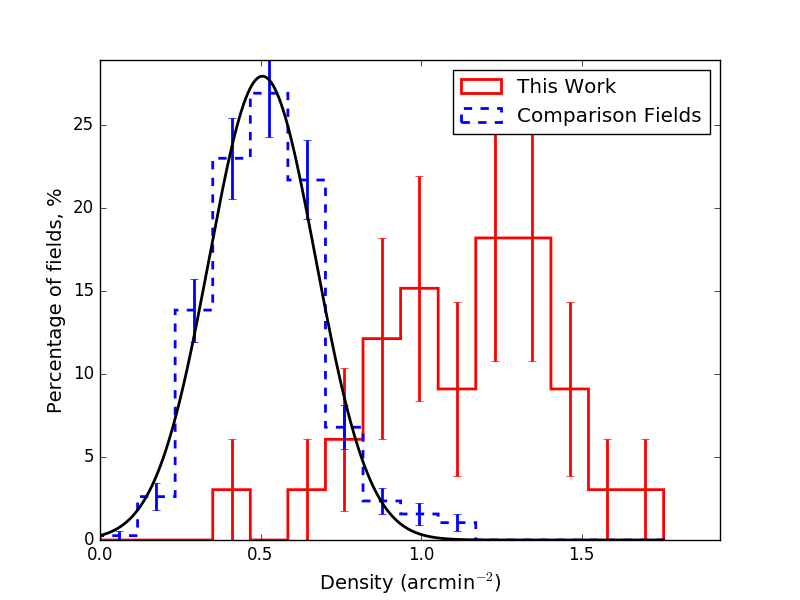}\\[-5ex]
\caption{Distribution of the number of galaxies with IRAC colour $\rm[3.6]-[4.5]>0.40$ per $\rm\sim26.2arcmin^{2}$ field. Radio-WISE selected fields are shown in red, comparison fields are indicated by dashed blue. A Gaussian fit has been added to the comparison fields to determine the level of overdensity for the radio-WISE selected fields, where the typical number of galaxies is $\rm0.47\pm0.16$.}
\label{rednumfig}
\end{figure}

We repeat the analysis in Section~\ref{papovichobj} for these redder-IRAC galaxies (see Fig.~\ref{rednumfig}). The results contrast drastically to the earlier results, suggesting that the radio-WISE selected galaxies inhabit significantly overdense regions of Group-II galaxies. The mean density for the comparison fields for the Group-II galaxies is $\rm0.47\pm0.16$ arcmin$\rm^{-2}$ (versus $\rm8.31\pm1.1$ arcmin$\rm^{-2}$ for Group-I in the comparison fields): $\rm97\%$ of the radio-WISE selected fields exhibit densities greater than the mean for the comparison fields and at the $\rm>1\sigma$ level; $\rm\sim76\%$ of fields are overdense by $\rm>3\sigma$ and $\rm 33\%$ by $>\rm5\sigma$. Since a large number of these fields exhibit significant overdensities with respect to the comparison fields, it is likely that the observed overdensity of Group-II galaxies is associated with the radio-WISE galaxy. These results suggest that the overdensity of Group-II galaxies, shown in Fig.~\ref{rednumfig}, is significant over all of the radio-WISE selected fields. Further, by repeating the K--S test for this distribution and comparing to the SpUDS and S-COSMOS fields, we find a K--S statistic of 0.83 with a $p$-value of 1.68$\rm\times10^{-19}$, and thus the samples are not drawn from the same underlying distribution.

\begin{figure} 
\includegraphics[trim={0.8cm 0cm 1.5cm 1.2cm},clip,width=\columnwidth]{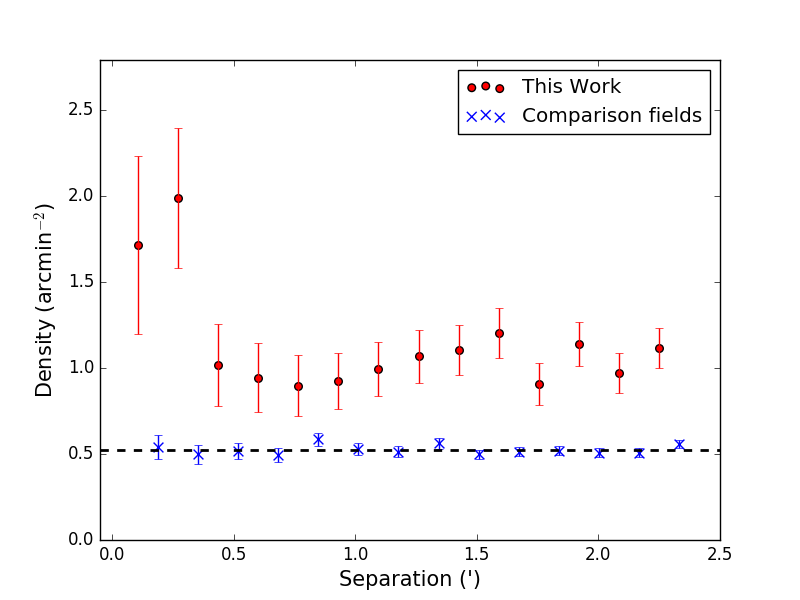}\\[-5ex]
\caption{The radial distribution of Group-II galaxies with IRAC colour $\rm[3.6]-[4.5]>0.4$ to a $\rm2.5\,'$ radius from the central radio-WISE galaxy. Radio-WISE selected fields are represented by red circles and comparison fields are indicated by blue crosses. The dashed line denotes the average density of the comparison fields for reference. See Fig.~\ref{papradialfig} for the radial distribution of Group-I galaxies for comparison.}
\label{redderradialfig}
\end{figure}

\begin{figure*}
\captionsetup[subfigure]{labelformat=empty}
\begin{tabular}{cc}
\subfloat[][]{\includegraphics[trim={0.34cm 0cm 1.45cm 1.2cm},clip,width=0.33\textwidth]{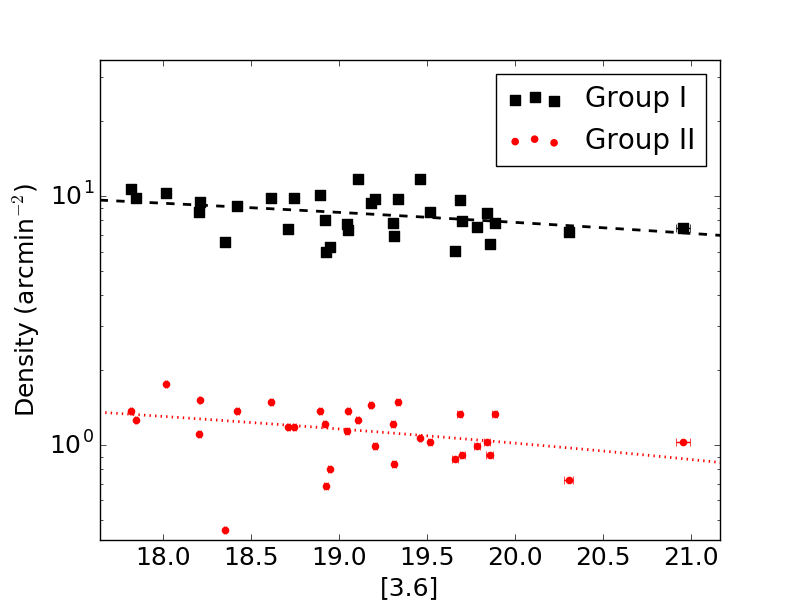}\label{3_6_vs_num}}
\subfloat[][]{\includegraphics[trim={0.34cm 0cm 1.45cm 1.2cm},clip,width=0.33\textwidth]{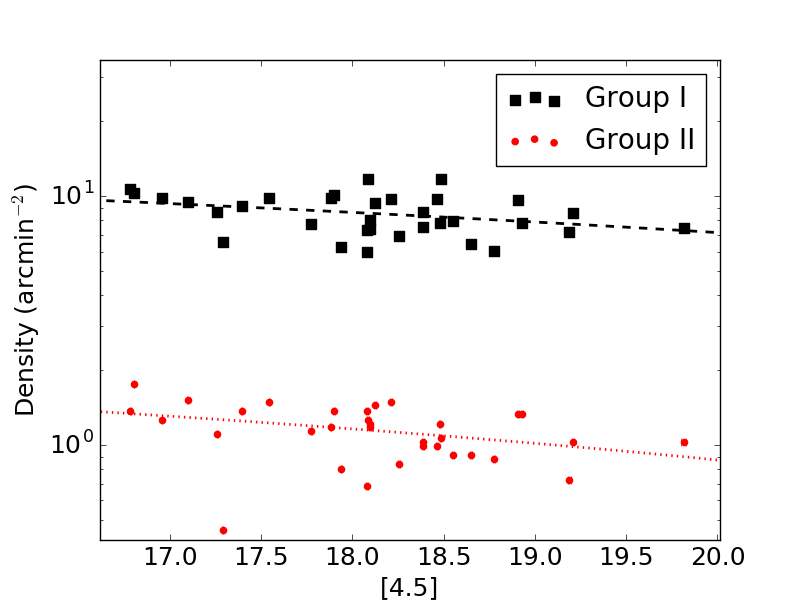}\label{4_5_vs_num}}
\subfloat[][]{\includegraphics[trim={0.34cm 0cm 1.45cm 1.2cm},clip,width=0.33\textwidth]{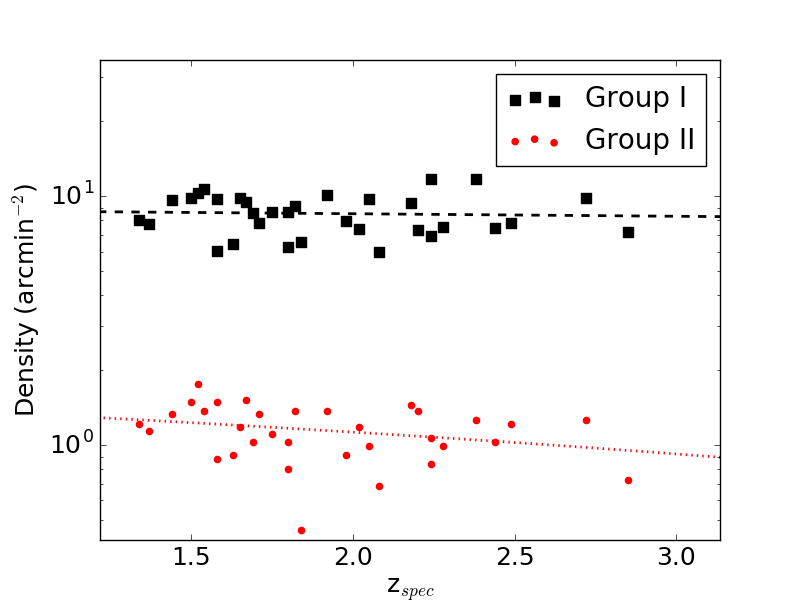}\label{z_vs_num}}\\[-6ex]
\end{tabular}
\caption{Comparison of the density of galaxies in the surrounding field, as a function of the features of the central radio-WISE selected galaxy, using the completeness limits in Section~\ref{completeness}. \protect\subref{3_6_vs_num} illustrates the distribution of [3.6] magnitude of the central radio-WISE selected galaxy against the density of the surrounding field. \protect\subref{4_5_vs_num} illustrates the distribution of [4.5] magnitude of the central radio-WISE selected galaxy against the density of the surrounding field. \protect\subref{z_vs_num} illustrates the distribution of redshift of the central galaxy with the density of the surrounding $\rm5.12\,'\times5.12\,'$ field. In each, black squares show the density of Group-I galaxies and red points illustrate the density of Group-II galaxies.}
\label{centralsfig}
\end{figure*}

To investigate the spatial distribution of the Group-II galaxies relative to the central radio-WISE galaxy, we stack the 33 fields to determine the radial distribution of these galaxies. As shown in Fig.~\ref{redderradialfig}, the overdensity of Group-II galaxies is fairly uniform across most of the field ($\rm\sim1.15\,$arcmin$\rm^{-2}$), generally $\rm\sim2$ times the average density found in the comparison fields, and less than the overdensity of $\rm\sim5$--6 found in \cite{SFJones15} ($\rm\sim0.38\,$arcmin$\rm^{-2}$) for rare SMGs around radio-WISE galaxies. Previous findings with ALMA around similar radio-WISE selected galaxies \citep{ASilva15} found a surface density of $\rm\sim19\,$arcmin$\rm^{-2}$ in the environment detected at $\rm870\,\mu$m, 10 times that of blank fields within $\rm10\,''$ of the central galaxy. Further, we find only a single match using the SMG positions presented in \cite{ASilva15}, albeit for a Group-I selected galaxy, suggesting that the Group-II galaxies are not IRAC counterparts to the SMGS in \cite{ASilva15}.

We find an increased density of Group-II galaxies within $\rm 0.5\,'$ of the radio-WISE galaxies with respect to comparison fields, suggesting angular clustering around the central galaxy on sub-arcmin scales, an effect that is not seen for brighter SMGs \citep{SFJones15}, which show no central concentration. This could be could be due to the $\rm15\,''$ beam size and the low density of detections per field in \cite{SFJones15}, however the overdensity of 4--6 in \cite{SFJones15} is greater than the $\rm\sim2\times$ overdensity shown in Fig.~\ref{redderradialfig}, where the S/N of the overdensity of Group-II galaxies is greater than 3. The Group-II galaxies in the $\rm0.5\,'$ peak are thus not IRAC counterparts of SMGs detected at $\rm850\,\mu$m. Further, using the positions of the SMGs in \cite{SFJones17}, we find that only one galaxy matches the positions of the SMGs around the radio-WISE selected galaxies, a Group-I galaxy, showing that these Group-II 
galaxies are not counterparts to the SMGs in \cite{SFJones15}. The SMGs in \cite{SFJones15, ASilva15} are likely luminous $\rm850\,\mu$m and 
$\rm850\,\mu$m galaxies respectively, and a different subset of galaxies associated with the radio-WISE selected galaxies than the IRAC selected Group-I and Group-II galaxies in this work.

The Group-II galaxies are likely to have similar redshifts to the radio-WISE galaxies. However, we see little evidence of angular clustering of galaxies in each individual radio-WISE selected field within $\rm0.5\,'$, implying that this is a statistical peak from the stacking of all 33 radio-WISE selected fields in Fig.~\ref{redderradialfig}. The result is expected, however, from the distribution of faint, redder [3.6] galaxies in the local environment (see Fig.~\ref{innercolorfig}). The statistical peak in Fig.~\ref{redderradialfig} could be composed of mainly fainter IRAC1 objects.

We now compare the density of the Group-I and Group-II galaxies in each field to determine if the density of galaxies in each subset is correlated. Using a Pearson correlation test, the correlation coefficient is 0.65 with a $p$-value of 4.17$\rm\times10^{-5}$, suggesting that we cannot accept the null hypothesis that these two datasets are correlated. Comparing the five densest and five sparsest fields in each subset, we find that only one field out of the five is in common between the densest fields and three out of the five are in common between the sparsest fields for both IRAC colour selections. This also suggests that there is little correlation between the density of the Group-I and Group-II galaxies, potentially owing to the Group-II galaxies having significantly higher redshifts or residing within a protocluster compared to the Group-I galaxies, given that galaxies become progressively redder with increasing redshift \citep{DStern05}.

Comparing the density of IRAC-selected galaxies expected to be at $z>$1.3 in the environment of the radio-WISE selected galaxies to the properties of the central target (Table~\ref{centrals}) in Fig.~\ref{centralsfig}. Using the Pearson's correlation test, we see little correlation between the density of Group-I galaxies and the redshift of the central radio-WISE galaxy, where the correlation coefficient is $\rm-0.05$ with a $p$-value of 0.79. The correlation between the mid-IR magnitudes of the targets and the density of the Group-I galaxies is -0.34 and -0.33 ($p$-value=0.05 and 0.06) for IRAC1 and IRAC2 respectively. This suggests a weak anti-correlation between the density of Group-I galaxies and the mid-IR magnitudes of the central radio-WISE selected galaxies.

We also compare the density of the $\rm[3.6]-[4.5]>0.4$ galaxies in each field with the properties of the central radio-WISE galaxy. Using the Pearson correlation test of the density and the redshift of the central radio-WISE galaxy, we find a coefficient of $\rm-0.29$ with a $p$-value of 0.11. This suggests we cannot rule out a weak anti-correlation between the density of Group-II galaxies and the redshift of the target. Comparing the magnitude of the central radio-WISE galaxy with density, we find a correlation coefficient of $\rm-0.36$ and $\rm-0.37$ ($p$-value=0.04 and 0.03) for IRAC1 and IRAC2 respectively. Given the $p$-values of this result, it is unlikely that the mid-IR flux densities of the radio-WISE galaxies are correlated with the density of their environment due to scatter. Overall, we find no correlation between the environment of the radio-WISE selected galaxies and the properties of the central radio-WISE galaxy, similar to findings by \cite{SFJones15}, although we cannot rule out a correlation between the redshift of the central target and the density of the field.

\subsection{Angular Auto-Correlation Function}
To further investigate the distribution of the two samples of IRAC-selected galaxies, we determine their two-point angular correlation function to investigate whether they are clustered. On relevant scales, this excess density usually takes the form of a power-law angular function.

\begin{equation}\label{limbereqn}
\omega(\theta) = \rm \bigg ( \frac{\theta}{\theta_{\: 0}} \bigg )^{1-\gamma}
\end{equation}
We used the Landy-Szalay estimator:

\begin{equation}\label{landyeqn}
\omega(\theta) = 1\: +\: \bigg(\frac{N_{r}}{N_{d}} \bigg)^{2} \frac{DD(\theta)}{RR(\theta)}\: - 2 \: \bigg(\frac{N_{r}}{N_d}\bigg) \frac{DR(\theta)}{RR(\theta)}
\end{equation}
\citep{SDLandy93} to determine the correlation function for galaxies in our dataset, and for similar populations in SpUDS and S-COSMOS. $N_{d}$ is the number of data points, $N_{r}$ is the number of random comparison points randomly projected onto the same area and used as a comparison to determine whether the real galaxies in the field are correlated. $ DD(\theta)$ is the number of pairs of data points with separation $\theta$, $RR(\theta)$ is the corresponding number of random pairs and $DR(\theta)$ is the number of data-random pairs. An unclustered sample should have $\rm\omega(\theta)\sim0$.

To avoid blended sources, the smallest separation we use in this work is 0.06$\rm\,'$ (3.6$\rm\,''$, $\rm\sim2\times$ PSF), also greater than the $\rm1.8\,''$ matching radius for galaxies between IRAC1 and IRAC2 in this survey. To further reduce the uncertainty, the full SpUDS and S-COSMOS fields are used, increasing the number of data pairs available for comparison. We used $\rm\sim14000$ galaxies in the Group-I magnitude and colour selection for the SpUDS field and the full S-COSMOS field. An additional source of uncertainty arises in the number of random objects used to compare to a randomly distributed sample. To reduce this uncertainty, we choose a bootstrap method \citep{astroML}, using 100 bootstraps for each field. Beyond this number of bootstraps, we see little improvement in the measurement error for the correlation for each radio-WISE field. Errors for each bin are calculated using the standard deviation of the correlation function for each bootstrap.

Previous angular two-point correlation functions for IRAC galaxies have found $\rm\theta_{0}\sim0.03\,'$ and $\rm\gamma\sim1.8$ \citep[e.g.][]{SOliver04, IWaddington07, Papovich08}. Typically, \cite{SOliver04} and \cite{IWaddington07} observe the clustering of galaxies at $z\sim0.75$ using SWIRE, with IRAC1 $\rm5\sigma$ flux density limits of $\rm3.7\,\mu$Jy and $\rm50\%$ completeness in IRAC1 at $\rm4\mu$Jy for \cite{SOliver04} and \cite{IWaddington07} respectively. \cite{Papovich08}, who introduced the selection technique in this work, selected $z>$1.3 galaxies using $\rm[3.6]-[4.5]>-0.1$, with $\rm5\sigma$ flux limits of 3.7 and $\rm5.4\mu$Jy for IRAC1 and IRAC2 respectively. It should be noted that, due to the size of the fields here, the angular correlation function for the radio-WISE selected fields is only probed on small scales of $\rm\lesssim5\,'$.

\begin{figure} 
\includegraphics[trim={0.8cm 0cm 2cm 0cm},clip,width=\columnwidth]{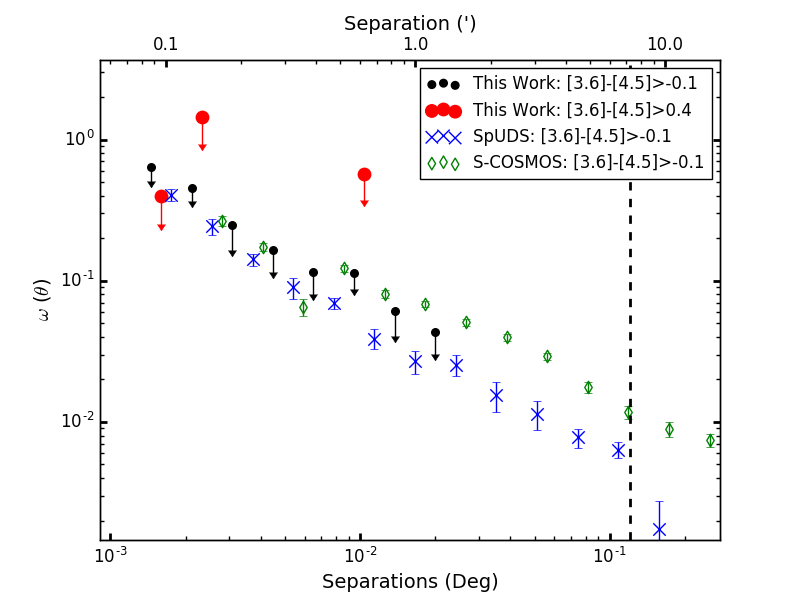}\\[-5ex]
\caption{Angular two-point correlation function for both colour selections in this work. $\rm2\sigma$ upper limits for the Group-I galaxies in the radio-WISE selected fields are shown by black circles, $\rm2\sigma$ upper limits for the Group-II galaxies in the radio-WISE selected fields given by red circles. The mean for the two comparison fields have been shown separately; the full 1$\rm\,deg^{2}$ SpUDS field is indicated by blue crosses, while the full 2$\rm\,deg^{2}$ S-COSMOS field is indicated by green diamonds. A line of best fit for the Group-I and Group-II radio-WISE fields are shown by a line and a red dashed line respectively. A vertical dashed line has been added to indicate the maximum separation attainable in the radio-WISE selected fields for reference.}
\label{Correlationallfig}
\end{figure}

The results of the two-point angular auto-correlation function for the radio-WISE selected fields are shown in Fig.~\ref{Correlationallfig}. For the Group-I galaxies, we find $\rm\theta_{0}$=$\rm0.026\,'\pm0.013\,'$ and $\gamma$=$\rm2.26\pm0.42$. We also list the number of galaxies used to create the correlation function in Table~\ref{autocorrbins}. The value of $\gamma$ for the radio-WISE selected fields is much larger than previous values observed in SWIRE \citep{SOliver04, IWaddington07}, suggesting that the clustering of galaxies decreases quicker than other fields. The value of $\rm\theta_{0}$ is roughly equal to the accepted value, and suggests that the Group-I galaxies in the radio-WISE selected fields are no more clustered than in blank fields. This is unusual, given that the galaxies selected in the SWIRE fields are generally at lower redshifts than this work and could be due to the large range of potential redshifts that the Group-I IRAC colour selection encompasses, such that galaxies at significantly different redshifts could be paired in this investigation. However, without deep coverage of these fields in other bands, determining photometric redshifts of these galaxies are unfeasible.

We see a slight difference in the two-point angular autocorrelation function between the SpUDS and S-COSMOS fields. This could be due to cosmic variance or the wide range of redshifts in the Group-I and Group-II colour selection. Using a K--S test for the comparison fields, we find a K--S statistic of 0.23 with a $p$-value of 0.83, suggesting that these two datasets are drawn from the same distribution. For the SpUDS field, using Group-I colour limits, we find $\rm\theta_{0}$=$\rm0.056\pm0.002\,'$ and $\gamma$=$\rm2.38\pm0.05$. For the S-COSMOS field, using the same colour selection, we find $\rm\theta_{0}$=$\rm0.029\,'\pm0.009\,'$ and $\gamma$=$\rm1.80\pm0.11$, in agreement with previous values for the two-point angular autocorrelation function. The value of $\gamma$ for the SpUDS field agrees with the value for the radio-WISE selected fields for the Group-I colour selection, while the value of $\gamma$ for the S-COSMOS field is smaller than the value for the radio-WISE selected fields, suggesting that the clustering of the Group-I galaxies in the radio-WISE selected fields decreases at the same rate or greater compared to blank fields for this colour selection. The value of $\rm\theta_{0}$ for the SpUDS field is greater than the radio-WISE selected fields for the Group-I selection, suggesting that the radio-WISE selected fields are no more clustered than blank fields for this IRAC colour selection.

We use a K--S test to determine whether there is a significant difference in the correlation function between the comparison fields and the radio-WISE selected fields. For the SpUDS field, we find a K--S statistic of 0.31 and a $p$-value of 0.65. Similarly, for the S-COSMOS field, we find a K--S statistic of 0.29 with a $p$-value of 0.73, suggesting that it is likely that the comparison fields and the radio-WISE selected fields are drawn from the same distribution. From these results, it is unlikely that many of the Group-I galaxies in the radio-WISE selected fields are associated with one another, although they may be related to the radio-WISE selected galaxy, given the peak in Fig.~\ref{papradialfig}. This lack of correlation is expected, given that the Group-I colour selection should only indicate the galaxies have a redshift $z>\rm1.3$, giving them a large range of potential redshifts. 

For the Group-II galaxies, we find $\rm\theta_{0}$=$\rm0.072\,'\pm0.013\,'$ and $\gamma$=$\rm3.10\pm0.20$. The value of $\gamma$ is significantly larger than the values for the Group-I selection suggesting that there is a significant decrease in clustering for these galaxies, although this is likely due to the lack of positive $\rm\omega(\theta)$ values. Given the lack of correlation at radii $\rm\lesssim0.5\,'$ for the Group-II galaxies in the radio-WISE fields, due to the small densities of galaxies at these scales, we are unable to comment on the correlation of the $\rm0.5\,'$ peak in Fig.~\ref{redderradialfig}. Due to the small densities of Group-II galaxies in the comparison fields, (see Fig.~\ref{rednumfig}), we combine the entire SpUDS and S-COSMOS fields. Using this method, we find $\rm\theta_{0}$=$\rm0.11\,'\pm0.04\,'$ and $\gamma$=$\rm2.27\pm0.22$. The values for the comparison fields are close to the values found in previous work \citep{SOliver04,IWaddington07}, suggesting that the Group-II selected galaxies in the comparison fields are more correlated than those in the radio-WISE selected fields. However, given the uncertainties associated with counting galaxies in the $\rm5.12\,'\times5.12\,'$ fields compared to the significantly reduced errors on larger scales, observations on larger fields around the radio-WISE galaxies would be needed to better understand the correlation for this IRAC colour selection.

\section{Discussion}\label{discussion}
In this section, we discuss the results of the radio-WISE selected fields with respect to the comparison fields and previous work on similar galaxies. It should be noted that we compare the overdensities of SMGs to IRAC-selected overdensities around radio-WISE galaxies and without additional information we cannot determine if galaxies in this work are IRAC counterparts of SMGs.

\subsection{Overdensities in Radio-WISE selected fields}
Targeted $Spitzer$ observations of fields containing radio-WISE galaxies suggest that these galaxies do not inhabit significantly overdense regions of Group-I ($\rm[3.6]-[4.5]>-0.1$) galaxies on scales extending over $\rm2.5\,'$. We also see no significant excess variance in the distribution of density of such galaxies in our fields.
 
Shallower observations ($\rm\sim150\,$s per field) of 90 Hot DOGs by \cite{RJAssef15} using $Spitzer$ with the same IRAC colour selection find a density of Group-I galaxies of $\rm\sim5$--6$\rm\,arcmin^{-2}$ within a $\rm95\%$ completeness limit of $\rm10\,\mu$Jy in IRAC2. Similar to findings by \cite{SFJones14} for 10 Hot DOGs at 850$\rm\,\mu$m, \cite{RJAssef15} suggest a lack of angular clustering around the central Hot DOG within $\rm2\,'$ radius. To compare the density of Group-I galaxies in our fields to the $Spitzer$ fields containing Hot DOGs in \cite{RJAssef15}, we cut our data at the same $\rm10\,\mu$Jy completeness limit in IRAC2. We find a density of 5$\,$arcmin$\rm^{-2}$, similar to the findings by \cite{RJAssef15}, albeit still with a central, statistical peak within $\rm\sim0.25\,'$ of the radio-WISE selected galaxy. This suggests that the density of Group-I galaxies in the radio-WISE selected fields is consistent with fields containing Hot DOGs, although there is an additional peak in the density towards the radio-WISE galaxies.

\subsection{Radial Distribution}
We see a $\rm>2\sigma$ peak ($\rm\sim1.85\,$arcmin$^{-2}$) in the density of Group-I galaxies within $\rm0.25\,'$ ($\rm\sim130\,$kpc) of 33 radio-WISE galaxies (Fig.~\ref{papradialfig}), a statistical finding. This peak implies an excess of associated galaxies on relatively small scales, consistent with findings for galaxies selected at longer wavelengths with ALMA \citep{ASilva15}. Beyond $\rm0.25\,'$, we see no excess density of Group-I galaxies out to $\rm2.5\,'$, showing that the central $\rm0.25\,'$ peak is responsible for the $\rm\sim10\%$ overdensity observed in Fig.~\ref{magnitudefig}. The central peak in density of Group-I galaxies ($\rm\sim10.7\,$arcmin$^{-2}$) is composed of fainter IRAC1 galaxies than the radio-WISE selected galaxies, in the magnitude range of 20.5$<$[3.6]$<$22, which do not generally exhibit redder-IRAC colours, as shown in Fig.~\ref{innercolor_vs_mag1}. These faint galaxies are unlikely to be artifacts of the image, as shown in Fig.~\ref{inset_d} we see that galaxies in this magnitude range are visibly discernible from the background of the image. These galaxies could be influenced by the central radio-WISE galaxy, although it is unclear what mechanisms in the environment or from the target galaxy would account for them. The lack of redder-IRAC colours suggests that the galaxies within $\rm0.25\,'$ of the central target are not heavily dust obscured and are unlikely to possess populations of older, redder stellar populations.

\cite{SFJones15} find an extended overdensity of very luminous sub-millimeter galaxies on $\rm1.5\,'$ radius scales, $\rm\sim4$--6 times that of blank fields. The very luminous SMGs in the environment of radio-WISE galaxies seen in \cite{SFJones15} are a different population to the Group-I galaxies in this work, given the different excess densities and lack of corresponding positions, where the overdensity in this work is $\rm\sim1.3$ compared to $\rm\sim4$--6 in \cite{SFJones15}. Despite the lower density of Group-I galaxies galaxies in this work, we see significantly greater angular clustering within $\rm0.25\,'$ of the radio-WISE selected galaxy (see Fig.~\ref{papradialfig}) than in the field. This could suggest that the Group-I galaxies are influenced by the central radio-WISE galaxy. A lack of star-forming galaxies in $\rm0.25\,'$ from the radio-WISE galaxy is observed in ALMA \citep{ASilva15}. However, it is unlikely that they are IRAC counterparts to the SMGs in \cite{ASilva15}, given the lack of corresponding positions and the difference in the observed excess between these works. Without further information on the galaxies in this peak, we cannot determine whether this is just a statistical fluctuation and how the radio-WISE galaxy could be influencing the galaxies within $\rm0.25\,'$ radius.

The peak in Fig.~\ref{papradialfig} is similar in spatial extent to the findings for HzRGs in \cite{DW13}, who suggested an association between these central HzRGs and the Group-I galaxies. The radio luminosities of the radio-WISE galaxies in this work are substantially lower than the HzRGS in CARLA, and have been selected for their less extended radio emission. The density in the peak is significantly less than that seen in CARLA, where $\rm\sim20$ detections arcmin$\rm^{-2}$ are found in the central $\rm0.25\,'$. From the luminosity functions in \cite{DW14}, the additional $\rm\sim0.5$ magnitude depth in CARLA does not double the expected density, so the radio-WISE selected fields are not as overdense in Group-I galaxies as the radio-loud galaxies in CARLA.

Unlike the Group-I galaxies, the Group-II galaxies in the radio-WISE selected fields show a significant spatially-uniform overdensity of 1.95 between $\rm0.5\,'-2.5\,'$ from the targets. This overdensity is roughly half the overdensity of SMGs in the environment of ultra-luminous, dusty galaxies in \cite{SFJones15}, which found overdensities of 4--6 for SMGs within $\rm1.5\,'$ of the radio-WISE galaxy compared to blank fields. These Group-II galaxies could be star-forming galaxies \citep{CMCasey16}, however IRAC is not sensitive to star formation at these redshifts, and any speculation is inconclusive without further information. Deeper, multi-wavelength observations are necessary to determine whether these Group-II galaxies are at the same redshift as the radio-WISE galaxy.

We see an additional peak in the distribution of the Group-II galaxies within $\rm0.5\,'$ from the central radio-WISE galaxy (Fig.~\ref{redderradialfig}). This peak represents an overdensity $\rm\sim2\sigma$ greater than the field, reaching 3--4 times the average density of the comparison fields, similar to the overdensity of Group-I galaxies in \cite{DW13} using similar completeness limits and for SMGs in \cite{SFJones15}. The overdensity of Group-II galaxies in this peak (3.53 times blank fields) is similar to the overdensity of 4--6 observed by \cite{SFJones15} on $\rm3\,'$ diameter fields, using shallower data than \cite{ASilva15}. This suggests that these Group-II galaxies are dust obscured similar to the SMGs in \cite{SFJones15}, although they are not IRAC counterparts. Given the smaller overdensities of these galaxies with respect to blank fields within $\rm0.5\,'$ of the central target, compared to the overdensities observed in \cite{ASilva15} ($\rm\sim10$ times blank fields), these Group-II galaxies are not IRAC counterparts to the ALMA detected SMGs around the same class of galaxy. The excess of Group-II galaxies is likely to be linked to the presence of the radio-WISE selected galaxy, given the $\rm>2\sigma$ statistical overdensities observed in the peak in Fig.~\ref{redderradialfig}.

We speculate that the excess of Group-II galaxies could be caused from infall into the cluster, producing bursts of star formation as the galaxies interact with gas from the Intra-Cluster Medium (ICM). Gas from the Inter-Stellar Medium (ISM) in these galaxies would be compressed by interacting with the ICM, and this dust-enshrouded star formation could produce the observed redder-IRAC colours. The Group-II galaxies in the $\rm0.5\,'$ peak are also likely to be dust obscured, star-forming galaxies similar to the redder galaxies in the Spiderweb field \citep{JDKurk04}. If Group-II galaxies $\rm\lesssim0.5\,'$ from the radio-WISE galaxy interacted with one another and/or the central radio-WISE galaxy in a potentially forming cluster, this could also stir up a dusty envelope to produce the observed redder IRAC colours \citep{HDannerbauer14}. Further work is necessary to understand the $\rm>2\sigma$ peak in the distribution of these red IRAC galaxies around the radio-WISE selected galaxy, and the spatially uniform excess density between $\rm0.5\,'-2.5\,'$ from the targets.

\section{Conclusions}
The results of $Spitzer$ IRAC imaging on $\rm5.12\,'\times5.12\,'$ fields centred on radio-WISE selected galaxies are:
\begin{itemize}
\item We find that the radio-WISE selected fields have a similar density of IRAC selected galaxies with colour $\rm[3.6]-[4.5]\rm>-0.1$ to blank fields, with only a modest $\rm10\%$ overdensity of Group-I selected galaxies. Using a K--S test for the density of Group-I galaxies in each $\rm5.12\,'\times5.12\,'$ field, we find a K--S statistic of 0.22 with a $p$-value of 0.08, suggesting that these fields are drawn from the same distribution and thus there is no significant overdensity of galaxies with IRAC colours indicating $z>$1.3 in the radio-WISE selected fields.
\item Using a redder IRAC colour selection of $\rm[3.6]-[4.5]>0.4$, we find a significant overdensity of Group-II galaxies by a factor of 1.95 on average with respect to blank fields within a radius of $\rm2.5\,'$ of the central radio-WISE galaxy. We see a significant overdensity of $\rm>3\sigma$ with respect to blank fields in $\rm76\%$ of the radio-WISE selected fields, suggesting that these galaxies inhabit dense regions of redder-IRAC galaxies.
\item We see smaller overdensities than \cite{SFJones15} and \cite{ASilva15} found for SMGs around galaxies also selected from WISE and radio surveys.
\item However, unlike the SMGs in \cite{SFJones15}, there is a statistical peak in density on scales $\rm<0.25\,'$ from the central radio-WISE galaxy for both colour selections. Many of these galaxies could have similar redshifts to the central radio-WISE galaxy.
\item The radio-WISE galaxies appear to be signposts for overdense regions of red galaxies \citep{ASilva15} and as such could be used to help understand the nature of galaxy and cluster formation during the epoch of peak star formation ($z\sim\rm2$) \citep{PRMEisenhardt12,SFJones15}.
\end{itemize}

With the launch of the James Webb Space Telescope (JWST)\footnote{https://www.jwst.nasa.gov/} in 2020, very deep multiband searches could be made to investigate the central peak of the Group-II galaxies in these fields, highlighted in Fig.~\ref{redderradialfig}. With the much improved depth of JWST observations, the peak from the faint galaxies can be further investigated to better understand the distribution and nature of these faint galaxies and uncover potentially further fainter counterparts. Deep multi-band searches would also be able to determine the extent to which foreground and background galaxies contribute to the peaks observed in the radial distribution of these fields with respect to the central radio-WISE galaxy.

\section*{Acknowledgements}
The authors wish to thank the staff in the Department of Physics and Astronomy at the Leicester University for their support. This research has made use of the NASA/IPAC Infrared Science Archive and the $Spitzer$ Space Telescope, which is operated by the Jet Propulsion Laboratory, California Institute of Technology, under contract with the National Aeronautics and Space Administration. Jordan Penney is supported by the Science and Technologies Facilities Council (STFC) studentship. We would also like to thank Omar Almaini and the COSMOS collaboration for making the UDS and COSMOS images and catalogues available. M. Kim was supported by the National Research Foundation of Korea (NRF), grant funded by the Korean government (MSIP) (No. 2017R1C1B2002879). R. J. Assef was supported by the FONDECYT grant (NO. 1151408). T.D.-S. acknowledges support from ALMA-CONICYT project 31130005 and FONDECYT regular project 1151239.




\bibliographystyle{mnras}
\bibliography{draftbib2} 



\appendix

\begin{table*}

\fontsize{9}{8}\selectfont

\caption{33 WISE-Selected Radio-Intermediate galaxies in the centre of the field, including the position, and photometry from the IRAC and WISE W3 and W4 bands with error. Galaxies marked with $\alpha$ by their designation can be found in \citep{SFJones15}, which details their expected IR luminosity that is not available in this work. For information on obtaining the spectroscopic redshifts for all radio-WISE selected galaxies, see previous work \citep{CJLonsdale15}. Error margins for IRAC bands are calculated using Source Extractor (see Section~\ref{SExtractor}). Positions, WISE magnitudes and errors were taken from the WISE All-Sky Catalogue.}

\centering

\begin{tabular}{ c c c c c c c c }

\hline \\

WISE Designation & RA (J2000) & DEC (J2000) & 3.6$\,\mu$m (mag) & 4.5$\,\mu$m (mag) & 12$\,\mu$m (mag) & 22$\,\mu$m (mag) & Redshift \\

\hline \\

W0304-3108 & 03:04:27.45 & -31:08:38.40 & 17.82 $\pm$ 0.005 & 16.78 $\pm$ 0.002 & 14.84 $\pm$ 0.04 & 13.81 $\pm$ 0.07 & 1.54 \\ \\

W0354-3308 & 03:54:48.22 & -33:08:27.24 & 19.04 $\pm$ 0.01 & 17.78 $\pm$ 0.003 & 15.17 $\pm$ 0.04 & 14.22 $\pm$ 0.13 & 1.37 \\ \\

W0519-0813 & 05:19:05.81 & -08:13:20.03 & 19.20 $\pm$ 0.01 & 18.47 $\pm$ 0.005 & 16.17 $\pm$ 0.13 & 14.79 $\pm$ 0.26 & 2.05 \\ \\

W0525-3614 & 05:25:33.50 & -36:14:40.92 & 19.84 $\pm$ 0.02 & 19.21 $\pm$ 0.01 & 16.08 $\pm$ 0.08 & 14.80 $\pm$ 0.23 & 1.69 \\ \\

W0526-3225 & 05:26:24.72 & -32:25:00.84 & 19.70 $\pm$ 0.01 & 18.55 $\pm$ 0.01 & 14.40 $\pm$ 0.03 & 12.83 $\pm$ 0.05 & 1.98 \\ \\

W0549-3739 & 05:49:30.07 & -37:39:39.95 & 19.89 $\pm$ 0.02 & 18.93 $\pm$ 0.006 & 16.22 $\pm$ 0.09 & 14.87 $\pm$ 0.23 & 1.71 \\ \\

W0613-3407 & 06:13:48.05 & -34:07:29.29 & 19.18 $\pm$ 0.01 & 18.13 $\pm$ 0.004 & 15.39 $\pm$ 0.05 & 14.19 $\pm$ 0.13 & 2.18 \\ \\

W0630-2121 & 06:30:27.82 & -21:20:58.92 & 19.69 $\pm$ 0.01 & 18.91 $\pm$ 0.01 & 15.37 $\pm$ 0.06 & 14.58 $\pm$ 0.23 & 1.44 \\ \\

W0642-2728 & 06:42:28.80 & -27:28:01.20 & 18.92 $\pm$ 0.01 & 18.10 $\pm$ 0.004 & 17.17 $\pm$ NA & 15.08 $\pm$ NA & 1.34 \\ \\

W0719-3349 & 07:19:12.72 & -33:49:44.77 & 19.86 $\pm$ 0.02 & 18.65 $\pm$ 0.01 & 15.70 $\pm$ 0.07 & 14.76 $\pm$ 0.25 & 1.63 \\ \\

W0729+6544 & 07:29:02.64 & 65:44:29.40 & 19.46 $\pm$ 0.005 & 18.48 $\pm$ 0.01 & 15.24 $\pm$ 0.05 & 13.77 $\pm$ 0.10 & 2.24 \\ \\

W0823-0624 & 08:23:11.28 & -06:24:08.42 & 18.21 $\pm$ 0.01 & 17.26 $\pm$ 0.003 & 14.96 $\pm$ 0.04 & 13.80 $\pm$ 0.10 & 1.75 \\ \\

W1308-3447 & 13:08:17.04 & -34:47:54.24 & 18.74 $\pm$ 0.01 & 17.88 $\pm$ 0.004 & 15.16 $\pm$ 0.05 & 13.96 $\pm$ 0.09 & 1.65 \\ \\

W1343-1136 & 13:43:31.44 & -11:36:09.72 & 19.31 $\pm$ 0.01 & 18.48 $\pm$ 0.01 & 15.90 $\pm$ 0.08 & 14.82 $\pm$ 0.22 & 2.49 \\ \\

W1400-2919 & 14:00:50.16 & -29:19:24.60 & 18.21 $\pm$ 0.002 & 17.10 $\pm$ 0.002 & 14.61 $\pm$ 0.03 & 13.67 $\pm$ 0.08 & 1.67 \\ \\

W1412-2020 & 14:12:43.20 & -20:20:11.04 & 18.42 $\pm$ 0.01 & 17.40 $\pm$ 0.003 & 15.22 $\pm$ 0.04 & 13.98 $\pm$ 0.11 & 1.82 \\ \\

W1434-0235 & 14:34:19.68 & -02:35:43.87 & 18.89 $\pm$ 0.01 & 17.90 $\pm$ 0.004 & 15.63 $\pm$ 0.06 & 14.67 $\pm$ 0.20 & 1.92 \\ \\

W1500-0649 & 15:00:48.72 & -06:49:39.83 & 18.61 $\pm$ 0.01 & 17.55 $\pm$ 0.003 & 14.47 $\pm$ 0.03 & 13.36 $\pm$ 0.07 & 1.50 \\ \\

W1513-2210 & 15:13:10.32 & -22:10:04.44 & 19.05 $\pm$ 0.01 & 18.08 $\pm$ 0.004 & 15.38 $\pm$ 0.07 & 13.82 $\pm$ 0.15 & 2.20 \\ \\

$\rm W1517+3523^{  \alpha}$ & 15:17:58.56 & 35:23:54.24 & 18.02 $\pm$ 0.01 & 16.80 $\pm$ 0.002 & 14.27 $\pm$ 0.03 & 13.11 $\pm$ 0.05 & 1.52 \\ \\

W1541-1144 & 15:41:41.76 & -11:44:09.24 & 19.34 $\pm$ 0.01 & 18.21 $\pm$ 0.005 & 15.34 $\pm$ 0.07 & 13.75 $\pm$ 0.11 & 1.58 \\ \\

W1634-1721 & 16:34:26.88 & -17:21:39.60 & 18.92 $\pm$ 0.01 & 18.08 $\pm$ 0.005 & 15.78 $\pm$ 0.10 & 14.62 $\pm$ 0.25 & 2.08 \\ \\

W1641-0548 & 16:41:07.20 & -05:48:26.86 & 18.35 $\pm$ 0.01 & 17.29 $\pm$ 0.003 & 15.25 $\pm$ 0.06 & 14.55 $\pm$ 0.2 & 1.84 \\ \\

W1653-0102 & 16:53:05.28 & -01:02:30.59 & 18.71 $\pm$ 0.01 & 18.10 $\pm$ 0.005 & 15.51 $\pm$ 0.08 & 14.64 $\pm$ 0.23 & 2.02 \\ \\

W1702-0811 & 17:02:04.56 & -08:11:07.41 & 20.31 $\pm$ 0.03 & 19.19 $\pm$ 0.01 & 15.40 $\pm$ 0.10 & 13.88 $\pm$ 0.18 & 2.85 \\ \\

W1703-0517 & 17:03:24.96 & -05:17:43.19 & 19.52 $\pm$ 0.01 & 18.39 $\pm$ 0.01 & 15.61 $\pm$ 0.12 & 13.86 $\pm$ 0.18 & 1.80 \\ \\

$\rm W1717+5313^{  \alpha}$ & 17:17:06.00 & 53:13:42.60 & 17.84 $\pm$ 0.005 & 16.96 $\pm$ 0.002 & 14.72 $\pm$ 0.03 & 13.79 $\pm$ 0.07 & 2.72 \\ \\

W1936-3354 & 19:36:22.56 & -33:54:20.52 & 19.31 $\pm$ 0.01 & 18.25 $\pm$ 0.005 & 15.50 $\pm$ 0.07 & 14.46 $\pm$ 0.20 & 2.24 \\ \\

W1951-0420 & 19:51:41.28 & -04:20:24.50 & 19.66 $\pm$ 0.02 & 18.78 $\pm$ 0.01 & 15.50 $\pm$ 0.08 & 14.05 $\pm$ 0.13 & 1.58 \\ \\

W1958-0746 & 19:58:01.68 & -07:46:09.30 & 18.95 $\pm$ 0.01 & 17.94 $\pm$ 0.004 & 15.25 $\pm$ 0.07 & 14.05 $\pm$ 0.12 & 1.80 \\ \\

W2000-2803 & 20:00:48.48 & -28:02:51.36 & 19.78 $\pm$ 0.02 & 18.39 $\pm$ 0.005 & 15.20 $\pm$ 0.07 & 14.24 $\pm$ 0.19 & 2.28 \\ \\

W2021-2611 & 20:21:48.00 & -26:12:00.00 & 20.96 $\pm$ 0.04 & 19.82 $\pm$ 0.01 & 16.40 $\pm$ 0.16 & 14.43 $\pm$ 0.19 & 2.44 \\ \\

W2059-3541 & 20:59:47.04 & -35:41:34.45 & 19.10 $\pm$ 0.01 & 18.08 $\pm$ 0.004 & 15.31 $\pm$ 0.06 & 14.60 $\pm$ 0.26 & 2.38 \\\hline

\end{tabular}

\label{centrals}

\end{table*}


\begin{table*}
\fontsize{9}{8}\selectfont
\caption{The number of data pairs ($DD(\theta)$ per separation shell used to compute the two-point angular autocorrelation function.}
\centering
\begin{tabular}{ c c c c c}
\hline \\
Separation Range (') & This Work: Group I & SpUDS: Group I & S-COSMOS: Group I & This Work: Group II \\
\hline \\
$\rm0.06\,'-0.09\,'$ & 1000  & 2960 & 11800 & 28 \\ \\[-1ex]
$\rm0.09\,'-0.13\,'$ & 2060 & 5450 & 33800 & 48 \\ \\[-1ex]
$\rm0.13\,'-0.18\,'$ & 3770 & 10500 & 66300 & 74 \\ \\[-1ex]
$\rm0.18\,'-0.27\,'$ & 7830 & 21500 & 127000 & 162 \\ \\[-1ex]
$\rm0.27\,'-0.39\,'$ & 15600 & 44700 & 284000 & 316 \\ \\[-1ex]
$\rm0.39\,'-0.57\,'$ & 31900 & 91800 & 580000 & 744 \\ \\[-1ex]
$\rm0.57\,'-0.83\,'$ & 61800 & 190000 & 1210000 & 1360 \\ \\[-1ex]
$\rm0.83\,'-1.20\,'$ & 119000 & 39400 & 2530000 & 2410 \\ \\[-1ex]
$\rm1.20\,'-1.75\,'$ & 215000 & 811000 & 5290000 & 4350 \\ \\[-1ex]
$\rm1.75\,'-2.54\,'$ & 361000 & 1660000 & 11000000 & 7580 \\ \\[-1ex]
$\rm2.54\,'-3.69\,'$ & 489000 & 3360000 & 22900000 & 8620 \\ \\[-1ex]
$\rm3.69\,'-5.38\,'$ & 363000 & 6700000 & 47100000 & 4640 \\ \\[-1ex]
$\rm5.38\,'-7.82\,'$ & 24000 & 12900000 & 95900000 & 148 \\ \\[-1ex]
$\rm7.82\,'-15.0\,'$ & 0 & 23700000 & 191000000 & 0 \\\hline
\end{tabular}
\label{autocorrbins}
\end{table*}


\bsp	
\label{lastpage}
\end{document}